\documentclass[preprint,floats,floatfix,tightenlines,11pt,eqsecnum,aps]{revtex4}
\usepackage{amsmath}
\usepackage{graphicx}
\begin{document}
\def\appls{\hbox{$<$\kern-.75em\lower 1.00ex\hbox{$\sim$}}}
\title{DETERMINATION OF $S$- AND $P$-WAVE HELICITY AMPLITUDES\\
AND NON-UNITARY EVOLUTION OF PION CREATION PROCESS\\
$\pi^- p \to \pi^- \pi^+ n$ ON POLARIZED TARGET}

\author{Miloslav Svec\footnote{electronic address: svec@hep.physics.mcgill.ca}}
\affiliation{Physics Department, Dawson College, Montreal, Quebec, Canada H3Z 1A4}
\date{September 12, 2007}

\begin{abstract}

We present the first model independent determination of $S$- and $P$-wave helicity amplitudes $A_n,n=0,1$, $A=S,L,U$ from CERN measurements of $\pi^- p \to \pi^- \pi^+ n$ on polarized target at small $t$ and dipion masses 580 -1080 MeV. The purely analytical determination of the helicity amplitudes is made possible by our finding analytical solutions for relative phase $\omega_{ij}=\Phi_{S_d}(j)-\Phi_{S_u}(i)$ between $S$-wave amplitudes of opposite transversities $u=up$ and $d=down$ for each set of the solutions of transversity amplitudes $A_u(i),A_d(j),i,j=1,2$. Of the six possible solutions for $\omega_{ij}$ only the solution with $\omega_{ij}=\pi$ yields physical helicity amplitudes $A_0(ij),A_1(ij),i,j=1,2$. Assigning $\rho^0(770)$ phase to the dominant $P$-wave helicity flip amplitude $L_1(ij)$ necessitates a phase of the $S$-wave helicity flip amplitude $S_1(ij)$ that is near to the $\rho^0(770)$ phase. Both amplitudes show resonant structures around 980 MeV for all solution sets $i,j=1,2$. These amplitudes are thus consistent with $\rho^0(770)-f_0(980)$ mixing observed previously in the reduced transversity amplitudes. The relative phases $\omega_{ij}=\pi$ satisfy certain self-consistency condition that must be satisfied in order for the four sets of solutions $A_u(i),A_d(j),i,j=1,2$ to be all physical amplitudes that can be identified with co-evolution amplitudes describing the interaction of the pion creation process with a quantum environment. This test on the phases $\omega_{ij}$ provides a new test of Kraus representation of the mixed final state density matrix in $\pi^- p \to \pi^- \pi^+ n$ and connects it to the experimentally measured amplitudes. This connection validates further the view of pion creation process $\pi^- p \to \pi^- \pi^+ n$ as an open quantum system interacting with a quantum environment. The probabilities $p_{ij}$ determining the final mixed state $\rho_f=\sum p_{ij}\rho_f(ij)$ in terms of solution states $\rho_f(ij)$ are experimentally measurable in measurements of recoil hyperon polarization in self-analyzing processes $\pi^- p \to \pi^- K^+ \Lambda^0$ and $K^- p \to \pi^- \pi^+ \Lambda^0$ on polarized targets. The probabilities $p_{ij}$ provide information about the quantum state of the environment.\\

\end{abstract}

\maketitle

\tableofcontents

\newpage
\section{Introduction.}

One of the fundamental assumptions of Quantum Theory is the unitary evolution of quantum systems. The assumption means that the quantum system $S$ is isolated and its future state can be calculated. Quantum Theory admits non-unitary evolution of open quantum systems that are not isolated but interact with a quantum environment $E$. In this case it is the system $S$ plus the environment $E$ that together undergo a unitary coevolution~\cite{kraus83,nielsen00,breuer02,bengtsson06}.\\

Theory of elementary particles is based on the assumption that the particle interaction is an isolated event in the Universe subject to unitary evolution~\cite{kaku93,itzykson05}. In 1982 Hawking questioned the universal validity of the unitarity assumption in particle interactions~\cite{hawking82,hawking84}. He suggested that particle interactions are processes that interact with an environment of quantum fluctuations of space-time metric which induce a non-unitary evolution of the particle interaction - at any energy. As the result of the non-unitary evolution pure initial states evolve into mixed final states. In contrast, a unitary evolution evolves pure initial states into pure final states.\\

Spin physics~\cite{leader01} is ideally suited to test Hawking's ideas. In our previous work~\cite{svec07a} we investigated the unitarity assumption in the pion creation process $\pi^- p \to \pi^- \pi^+ n$ measured at CERN on polarized target at 17.2 GeV/c~\cite{lutz78,becker79a,becker79b,chabaud83,rybicki85}. Following the initial work by Lutz and Rybicki~\cite{lutz78} we used a spin formalism to relate final state density matrix to transversity amplitudes with definite dipion spin, helicity and naturality. Imposing the purity of the final states for specific initial pure states as required by the unitarity leads to a set of constraints on the transversity amplitudes. The unitarity constraints are violated by measured amplitudes at large momentum transfers t obtained in a model independent amplitude analysis of CERN data~\cite{rybicki85}. This result suggests that in $\pi^- p \to \pi^- \pi^+ n$ pure states evolve into mixed states and that the pion creation process behaves as an open quantum system interacting with a quantum environment.\\

There are other indications of a non-unitary evolution in $\pi^- p \to \pi^- \pi^+ n$. At any dipion mass $m$, the $S$- and $P$-wave subsystem of reduced density matrix measured at  polarized target is analytically solvable in terms of reduced $S$-and $P$-wave transversity amplitudes with definite $t$-channel naturality~\cite{svec07b}. There are two physical solutions $A_u(i),i=1,2$ and $A_d(j),j=1,2$ for transversity amplitudes with target nucleon transversity "up" and "down", respectively~\cite{svec07b}. Associated with each set of the amplitudes is a final state density matrix $\rho_f(ij)$. Each $\rho_f(ij)$ carries specific information about the pion creation process and selecting one of the four sets of solutions amounts to a loss of such information. We can retain all the information if we assume that the measured final state density matrix is a mixed state of the solutions $\rho_f(ij)$
\begin{equation}
\rho_f=p_{11}\rho_f(11)+p_{12}\rho_f(12)+p_{21}\rho_f(21)+p_{22}\rho_f(22)
\end{equation}
where the probabilities $\sum p_{ij}=1$. In this case the pure initial states will evolve into mixed final states even when $\rho_f(ij)$ are pure states.\\

Observation of $\rho^0(770)-f_0(980)$ mixing in $\pi^- p \to \pi^- \pi^+ n$ provides another evidence for non-unitary dynamics of this process. Previous amplitude analyses of CERN data at small and large t found a rho-like resonance $\sigma(770)$ in the $S$-wave 
amplitudes~\cite{becker79b,rybicki85,svec96,svec97a,svec02a}. In a new high resolution amplitude analysis of CERN data at small t~\cite{svec07b} we show that this resonance must be $\rho^0(770)$ and that data indicate $\rho^0(770)-f_0(980)$ mixing also in the $P$-waves. This interpretation explains why no rho-like resonance has been observed in $S$-wave amplitudes in $\pi^- p \to \pi^0 \pi^0 n$ where $P$-waves do not contribute~\cite{gunter01}. In Ref.~\cite{svec07b} we propose to account for the $\rho^0(770)-f_0(980)$ mixing by introducing a new $CPT$ violating interaction of the produced resonant $q \overline{q}$ modes with interacting degrees of freedom of the environment which are identified with quantum numbers labeling the solutions for the amplitudes.\\

In Section II. we show that the requirement (1.1) that all four sets of solutions for transversity amplitudes $A_u(i),A_d(j),i,j,=1,2$ are accepted physical solutions leads to a self-consistency condition among relative phases $\omega_{ij}=\Phi_{S_d}(j)-\Phi_{S_u}(i)$. In Section III. we introduce Kraus representation of the mixed final state density matrix (1.1) arising from a unitary co-evolution of the pion creation process with a quantum environment. We define co-evolution amplitudes which are identified with the measured
solutions for transversity amplitudes. The co-evolution amplitudes must satisfy the self-consistency condition so that its experimental test is a test of validity of Kraus representation in pion creation process $\pi^- p \to \pi^- \pi^+ n$.\\

It has been generally believed that the determination of the relative phase $\omega$ requires difficult measurements of recoil nucleon polarization which involve the required bilinear terms between the amplitudes with nucleon transversities "up" and "down". In this work we show that the phases $\omega_{ij}$ can be determined from measurements on transversely polarized targets using relationships between helicity and transversity amplitudes.  In Section IV. we present the helicity amplitudes $A_n(ij),n=0,1$ and the expressions for their moduli and bilinear terms in terms of known bilinear terms of measured reduced transversity amplitudes and unknown $\cos \omega$ and $\sin \omega$. In Section V. we impose a self-consistency requirement between the product of two moduli squared and the square of magnitude of the corresponding bilinear term. This trigonometric condition yields 6 solutions for the pairs $\cos \omega$ and $\sin \omega$ for each solution set $i,j=1,2$.\\

The results are presented in Section VI..Two solutions with $\cos \omega \neq 0$ and $\sin \omega \neq 0$ are rejected because the corresponding solutions for helicity amplitudes exhibit unphysical chaotic behaviours and lack a clear resonant structure at $\rho^0(770)$ mass in $P$-wave amplitudes. Three solutions with $\cos \omega=0$, $\sin \omega = \pm 1$ and $\cos \omega =+1$, $\sin \omega =0$ are also rejected because the corresponding helicity amplitudes do not satisfy the requirement of pion exchange dominance of helicity flip amplitudes. The remaining solution with $\cos \omega=-1$, or $\omega = \pi$, yields a unique solution for helicity amplitudes that satisfy the requirement of pion exchange dominance and shows $\rho^0$ resonant structures in the $P$-wave. Assigning $\rho^0(770)$ phase to the dominant $P$-wave helicity flip amplitude $L_1(ij)$ necessitates a phase of the $S$-wave helicity flip amplitude $S_1(ij)$ that is near to the $\rho^0(770)$ phase. Both amplitudes show resonant structures around 980 MeV for all $i,j=1,2$. These amplitudes are thus consistent with $\rho^0(770)-f_0(980)$ mixing observed previously in the reduced transversity amplitudes~\cite{svec07b}.\\

The physical solutions for the relative phases $\omega_{ij}=\pi$ satisfy a self-consistency condition that must be satisfied by co-evolution amplitudes. The transversity amplitudes $A_u(i),A_d(j),i,j=1,2$ can thus be identified with co-evolution amplitudes connecting the Kraus representation to the experimentally measured amplitudes. This connection validates further the view of pion creation process $\pi^- p \to \pi^- \pi^+ n$ as an open quantum system interacting with a quantum environment~\cite{svec07a}.\\

In Section VII. we show explicitely that the unique solution for $\omega_{ij}=\pi$ leads to a unique solution for the complete set of $S$- and $P$-wave density matrix elements defining each state $\rho_f(ij)$ which results in a unique form of Kraus representation of the mixed final state (1.1). In Section VIII. we show how measurements of recoil hyperon polarization in $K^- p \to \pi^- \pi^+ \Lambda^0$ and $\pi^- p \to \pi^- K^+ \Lambda^0$ on polarized target can be used to determine the probabilities $p_{ij}$ in the mixed final state (1.1). The probabilities $p_{ij}$ provide information about the quantum state of the environment. The paper closes with a summary in Section IX..

\newpage
\section{$S$- and $P$-wave amplitude analysis of $\pi^- p \to \pi^- \pi^+ n$ on transversely polarized target and self-consistency of the central hypothesis.}

Amplitude analysis is an integral final part of any measurement of $\pi^- p \to \pi^- \pi^+ n$ on transversely polarized target when recoil nucleon polarization is not observed. It is a model independent conversion of the measured reduced density matrix elements into moduli and phases of nucleon transversity amplitudes with definite dipion spin $J$, helicity $\lambda$ and $t$-channel naturality $\eta$~\cite{svec07b}. For any dipion mass $m$ the $S$- and $P$-wave subsystem of the reduced density mantrix is analytically solvable in terms of reduced transversity amplitudes $A$ and $\overline{A}$ with target nucleon transversity $\tau=u$ (up) and $\tau=d$ (down), respectively~\cite{svec07b}. In our notation~\cite{svec07b} $A=S,L$ are unnatural exchange $S$- and $P$-wave amplitudes with helicity $\lambda=0$ while $A=U,N$ are unatural and natural exchange $P$-wave amplitudes corresponding to combinations of $\lambda=\pm 1$~\cite{svec07a}.\\

In a previous work~\cite{svec07b} we have shown that amplitude analysis of the complete $S$- and $P$-wave subsystem of the reduced density matrix yields four sets of solutions $A(i),N(ij),\overline{A}(j),\overline{N}(ij),i,j=1,2$ where $A=S,L,U$. This requires measurements with target polarization not only with transverse component $P_y$ but also with planar components $P_x$ and $P_z$. Such measurements are feasible with modern polarized targets~\cite{leader01}.\\

For any set of solutions $i,j=1,2$ the reduced transversity amplitudes $A,\overline{A}$ are defined in terms of transversity amplitudes $A_u(i),A_d(j)$ as follows~\cite{svec07b} 
\begin{equation}
S(i)=|S_u(i)|, \quad \overline {S}(j) = |S_d(j)|
\end{equation}
\[
L(i)=|L_u(i)| \exp i \left (\Phi_{L_u}(i)-\Phi_{S_u}(i) \right), \quad
\overline {L}(j)=|L_d(j)| \exp i \left ( \Phi_{L_d}(j)-\Phi_{S_d}(j) \right )
\]
\[
U(i)=|U_u(i)| \exp i \left (\Phi_{U_u}(i)-\Phi_{S_u}(i) \right). \quad
\overline {U}(j)=|U_d(j)| \exp i \left ( \Phi_{U_d}(j)-\Phi_{S_d}(j) \right )
\]
\[
N(ij)=|N_u(i)| \exp i \left (\Phi_{N_u}(i)-\Phi_{S_d}(j) \right), \quad
\overline {N}(ij)=|N_d(j)| \exp \left ( i\Phi_{N_d}(j)-\Phi_{S_u}(i) \right )
\]
where $\Phi_{A_\tau}$ is the phase of the amplitude $A_\tau$. The reduced transversity amplitudes are related to transversity amplitudes by phase factors
\begin{equation}
A_u(i)=A(i) \exp {i \Phi_{S_u}(i)}, \qquad A_d(j)=\overline {A}(j) \exp {i \omega_{ij}} \exp {i \Phi_{S_u}(i)}
\end{equation}
for unnatural exchange amplitudes $A=S,L,U$ and  
\begin{equation}
N_u(i)=N(ij) \exp {i \omega_{ij}} \exp {i \Phi_{S_u}(i)}, \qquad N_d(j)=\overline {N}(ij) \exp {i \Phi_{S_u}(i)}
\end{equation}
for natural exchange amplitude $N$. In (2.2) and (2.3) $\Phi_{S_u}(i)$ is the arbitrary absolute phase and
\begin{equation}
\omega_{ij}=\Phi_{S_d}(j)-\Phi_{S_u}(i)
\end{equation}
is the relative phase between $S$-wave amplitudes of opposite transversity. The moduli $|N(ij)|=|N_u(i)|$ and $|\overline{N}(ij)|=|N_d(j)$ depend only on the indices $i$ and $j$, respectvely, and are determined by measurements on transversely polarized target. In contrast, the phases of  
$N(ij)$ and $\overline{N}(ij)$ depend on both indices and are determined by measurements with planar components of target polarization.\\

Up to now it has been generally believed that experimental determination of the relative phases $\omega_{ij}$ requires difficult measurements of planar components of recoil nucleon polarization which involve the interferences between transversity amplitudes $A_u$ and $A_d$~\cite{lutz78,becker79a,becker79b,svec96,svec97a}. Each set $i,j=1,2$ of reduced transversity amplitudes would determine the corresponding phase $\omega_{ij}$. In Section IV. of this work we shall show that the phases $\omega_{ij}$ can be determined analytically in a process of converting the reduced $S$- and $P$-wave transversity amplitudes $A$ and $\overline{A}$, $A=S,L,U$ into $S$- and $P$-wave helicity amplitudes $A_n$ with definite $t$-channel naturality where $n=0,1$ is nucleon helicity flip. This process also involves the necessary interferences between transversity amplitudes $A_u$ and $A_d$.\\

Our central hypothesis (1.1) that all sets of transversity amplitudes $A_u(i),A_d(j),i,j=1,2$ are valid physical solutions imposes experimentally testable self-consistency constraint on the relative phases $\omega_{ij}$. Taken separately, each set $A_u(i),A_d(j)$ has its own absolute phase $\Phi_{S_u}(i)$ and $\omega_{ij}$ are in general not related. Taken together, all sets $A_u(i),A_d(j),i,j=1,2$ can have only one absolute phase, say $\Phi_{S_u}(1)$. Then we can write for $A=S,L,U$
\begin{eqnarray}
A_u(1) & = & \exp(i\Phi_{S_u}(1))A(1)\\
\nonumber
A_u(2) & = & \exp(i\Phi_{S_u}(1)) \exp(i\xi)A(2)\\
\nonumber
A_d(1) & = & \exp(i\Phi_{S_u}(1))\exp(i\omega_{11})\overline{A}(1)= \exp(i\Phi_{S_u}(1))
\exp(i\xi)\exp(i\omega_{21})\overline{A}(1)\\
\nonumber
A_d(2) & = & \exp(i\Phi_{S_u}(1))\exp(i\omega_{12})\overline{A}(2)= \exp(i\Phi_{S_u}(1))
\exp(i\xi)\exp(i\omega_{22})\overline{A}(2)
\end{eqnarray}
where 
\begin{equation}
\xi=\Phi_{S_u}(2)-\Phi_{S_u}(1)
\end{equation}
From the definitions $\omega_{11}=\Phi_{S_d}(1)-\Phi_{S_u}(1)$ and $\omega_{21}=\Phi_{S_d}(1)-\Phi_{S_u}(2)$ it follows that $\omega_{11}-\omega_{21}=\xi$ while from the definitions $\omega_{12}=\Phi_{S_d}(2)-\Phi_{S_u}(1)$ and $\omega_{22}=\Phi_{S_d}(2)-\Phi_{S_u}(2)$ it follows that $\omega_{12}-\omega_{22}=\xi$. The phases $\omega_{ij}$ are not independent but satisfy a condition
\begin{equation}
\xi=\omega_{11}-\omega_{21}=\omega_{12}-\omega_{22}
\end{equation}
For each set of reduced transversity amplitudes $A(i),\overline{A}(j)$, $A=S,L,U$ the conversion to helicity amplitudes provides an independent determination of the corresponding phase $\omega_{ij}$ for the set. If all the transversity amplitudes $A_u(i),A_d(j),i,j=1,2$ form a self-cosistent set of valid physical solutions then the condition (2.7) must be satisfied. Any violation of the condition (2.7) means that the calculation of $\omega_{11}$ and $\omega_{21}$ require phases $\Phi_{S_u}(1)$ and $\Phi_{S_u}(2)$ different from those in the calculation of $\omega_{12}$ and $\omega_{22}$.  In such a case the four solutions for the transversity amplitudes $A_u(i),A_d(j)$ do not form a self-consistent set of valid physical solutions. The condition (2.7) is thus an essential test of the central hypothesis (1.1).

\section{Kraus representation and its test in $\pi^- p \to \pi^- \pi^+ n$.}

In order to introduce the central concept of co-evolution amplitudes we first briefly review the Kraus representation for for reduced density matrices of open quantum systems interacting with an environment. It is the co-evolution amplitudes which involve the interacting degrees of freedom of the environment and with which we shall identify the solutions for the transversity amplitudes.\\

The co-evolution of an open quantum system $S$ with a quantum environment $E$ is a unitary evolution~\cite{kraus83,nielsen00}
\begin{equation}
\rho_f(S,E)=U \rho_i(S,E)U^+ =U \rho_i(S) \otimes \rho_i(E) U^+
\end{equation}
The initial state of the environment is in general a mixed state
\begin{equation}
\rho_i(E)= \sum \limits_{\ell} p_{\ell \ell'} |e_\ell><e_{\ell'}|
\end{equation}
where $|e_\ell>$ are quantum states of interacting degrees of the environment and $\sum \limits_\ell p_{\ell \ell}=1$. The Hilbert space of the environment has a finite dimension. It is given by a condition $\dim {H(E)} \leq \dim{H_i(S)} \dim{H_f(S)}$~\cite{nielsen00}. Here $H(E)$, $H_i(S)$ and $H_f(S)$ are Hilbert spaces of the environment $E$ and of the initial and final states of the system $S$. After the interaction the system $S$ is fully described by reduced density matrix $\rho_f(S)$ given by Kraus representation
\begin{equation}
\rho_f(S)=Tr_E(\rho_f(S,E))= \sum \limits_\ell \sum \limits_{m,n} 
p_{mn} S_{\ell m} \rho_i(S) S_{n \ell}^+
\end{equation}
where the operators $S_{\ell m}=<e_\ell|U|e_m>$ satisfy a completness relation 
$\sum \limits_\ell \sum \limits_{m,n} S_{n \ell}^+ S_{\ell m}=I$. Kraus representation assumes that the initial states $\rho_i(S)$ and $\rho_i(E)$ are separable to ensure complete positivity of $\rho_f(S)$~\cite{bengtsson06}.\\

In our next step we associate the two solutions for transversity amplitudes $A_u(i)$ and $A_d(j)$, $i,j=1,2$ with two single qubit states $|i>$ and $|j>$, respectively. Then the hypohesis (1.1) allows us to identify the four degrees of freedom of the environment $|e_\ell>$ allowed by the condition  
$\dim {H(E)} \leq \dim{H_i(S)} \dim{H_f(S)}=(2s_p+1)(2s_n+1)=4$ with the four two-qubit states  $|e_\ell>=|i>|j>$. The co-evolution amplitudes are then defined by matrix elements
\begin{equation}
U^{J \eta}_{\lambda, \tau}(\ell m)=<J \lambda \eta,\tau_n|<e_\ell|U|e_m>|0 \tau_p>
\end{equation}
where $J$, $\lambda$ and $\eta$ are dipion spin, helicity and $t$-channel naturality, and $\tau \equiv \tau_p$ and $\tau_n$ are transversities of target proton and recoil neutron, respectively. Since the transversity amplitudes can possess only one solution at a time, the co-evolution amplitudes 
$U^{J \eta}_{\lambda, \tau}(\ell m)$ must be diagonal for any dipion spin $J$ and naturality $\eta$
\begin{equation}
U^{J\eta}_{\lambda, \tau}(\ell m)=U^{J\eta}_{\lambda, \tau}(\ell \ell)\delta_{\ell m}=
U^{J\eta}_{\lambda, \tau}(ij,ij) \delta_{ij,i'j'} \equiv
U^{J\eta}_{\lambda, \tau}(ij) \delta_{ij,i'j'} 
\end{equation}
where
\begin{equation}
U^{J\eta}_{\lambda, u}(ij)=
<J\lambda \eta,\tau_n|<ij|U|ij>|0 u>
\end{equation}
\[
U^{J\eta}_{\lambda, d}(ij)=
<J\lambda \eta,\tau_n|<ij|U|ij>|0 d>
\]

The requirement that the Kraus representation leaves invariant the spin formalism used in the data analysis necessitates~\cite{svec07a} that the co-evolution amplitudes $U^{J \eta}_{\lambda, \tau}(ij)$ transform under $P$-parity as a two-body $P$-parity conserving process $\pi^- + p \to "J(\pi^- \pi^+)" + n$ with parity $P=(-1)^J$ for the dipion states $"J(\pi^- \pi^+)"$~\cite{svec07a}. This means that there is no vector associated with the quantum states $|i>|j>$ of the environment. In particular, there is no energy-momentum associated with these quantum states. The interacting hadrons conserve their energy-momentum and there is no exchange of energy-momentum with the environment, in agreement with the original proposal by Hawking for particle processes interacting with quantum fluctuations of the space-time metric~\cite{hawking82}. Instead, the interaction with the environment is a non-dissipative dephasing process. The co-evolution amplitudes $U_{fi}$ can then be written in a form
\begin{equation}
U_{fi}=I_{fi}+i(2 \pi)^4 \delta^4(P_f-P_i)T_{fi}
\end{equation}
where $P_i$ and $P_f$ are total four-momenta of the initial and final hadron states and $T_{fi}$ is the transition matrix for the process $|\pi^-p>+|i>|j> \to |\pi^- \pi^+ n>+|i>|j>$.\\ 

In a final step we identify the co-evolution transtion amplitudes with solutions for transversity amplitudes of any dipion spin
\begin{equation}
T^{J\eta}_{\lambda, u}(ij)=
<J\lambda \eta,\tau_n|<ij|T|ij>|0 u>=A^{J\eta}_{\lambda, u}(i)
\end{equation}
\[
T^{J\eta}_{\lambda, d}(ij)=
<J\lambda \eta,\tau_n|<ij|T|ij>|0 d>=A^{J\eta}_{\lambda, d}(j)
\]
where $A^{J\eta}_{\lambda, u}(i)$, $A^{J\eta}_{\lambda, d}(j)$, $i,j=1,2$ are the experimental solutions for transversity amplitudes. The set of co-evolution amplitudes $T^{J\eta}_{\lambda, \tau}(ij)$ is a consitent set with a single absolute phase $\Phi^{0-1}_{0,u}(11)$ that satisfies the consistency conditions (2.7). The consistent identification of the co-evolution amplitudes with the experimentally determined solutions for transversity amplitudes requires that the latter also satisfy the conditions (2.7) so that $\Phi^{0-1}_{0,u}(11)=\Phi_{S_u}(1)$. Since we have identified the interacting degrees of the environment with the quantum numbers labelling the solutions for transversity amplitudes, the test of condition (2.7) is de facto a test of Kraus representation of the final state $\rho_f(S)$ in pion creation process $\pi^- p \to \pi^- \pi^+ n$.\\

In our previous work~\cite{svec07b} we have introduced a quantum number $g=\pm1$ to label the solutions for the amplitudes. Instead of using the solution qubits $|i>$ and $|j>$ to define the quantum states $|i>|j>$ of the environment and the co-evolution amplitudes $T^{J \eta}_{\lambda, \tau}(ij)$, we could have used the qubits $|g_u>$ and $|g_d>$ to define equivalent states $|g_u>|g_d>$ of the environment and the equivalent co-evolution amplitudes $T^{J \eta}_{\lambda, \tau} (g_u g_d) \equiv A^{J \eta}_{\lambda, \tau}(g_\tau)$. These states more closely reflect the qubit nature of the interacting degrees of the environment and may possess a deeper physical meaning. However, in the context of the present work we prefer to use the labels $|i>$ and $|j>$ for the solutions.

\section{Helicity amplitudes and their bilinear terms.}

Helicity amplitudes with definite $t$-channel naturality were defined in 
Ref.~\cite{lutz78,svec07a} for any dipion spin $J$ and helicity $\lambda$ and their relations to transversity amplitudes of definite $t$-channel naturality were given. The $S$- and $P$-wave helicity nonflip and flip amplitudes $A_0$ and $A_1$ are related to transversity amplitudes $A_\tau, \tau=u,d$, $A=S,L,U,N$ by relations
\begin{equation}
A_n={(-i)^n \over{\sqrt{2}}}(A_u+(-1)^n A_d)
\end{equation}
where $n=0,1$. In terms of reduced transversity amplitudes we can write for the unnatural exchange amplitudes $A=S,L,U$
\begin{equation}
A_n={(-i)^n \over{\sqrt{2}}}(A+(-1)^n \overline{A}\exp(i \omega))\exp(i\Phi_{S_u})
\end{equation}
where $\omega=\Phi_{S_d}-\Phi_{S_u}$.\\

Measurements on transversely polarized target yield information on reduced density matrix elements $\rho^0_u$ and $\rho^0_y$. Amplitude analysis of this data results in four sets of solutions for the moduli of reduced transversity amplitudes $|A(i)|,|\overline{A}(j)|,i,j=1,2$. Each set comes with a fourfold sign ambiguity in the phases of the amplitudes $(A(i),A(j))_{\epsilon \overline{\epsilon}}$ where 
$\epsilon \overline{\epsilon}=++,+-,-+,--$ are the signs of relative phases $\Phi_{LS}(i)=\Phi_{L_u}(i)-\Phi_{S_u}(i)$ and $\overline{\Phi}_{LS}(j)=\Phi_{L_d}(j)-\Phi_{S_d}(j)$. These phases do not change sign as a function of dimeson mass $m$ in any of the 8 reactions analysed in Ref.~\cite{svec07b}. The change of sign of $\Phi_{LS}(i)$ ($\overline{\Phi}_{LS}(j)$) which is the phase af the amplitude $L(i)$ ($\overline{L}(j)$) results in the change of sign of the phase of amplitude $U(i)$ ($\overline{U}(j)$)~\cite{svec07b}, or complex conjugation of amplitudes $A(i)$ ($\overline{A}(j)$). We can thus write the four sets of phases for each $i,j$
\begin{eqnarray}
(A(i),\overline{A}(j))_{++} & = & (A(i),\overline{A}(j))\\
\nonumber
(A(i),\overline{A}(j))_{+-} & = & (A(i),\overline{A}(j)^*)\\
\nonumber
(A(i),\overline{A}(j))_{-+} & = & (A(i)^*,\overline{A}(j))=((A(i),\overline{A}(j))_{+-})^*\\
\nonumber
(A(i),\overline{A}(j))_{--} & = & (A(i)^*,\overline{A}(j)^*)=((A(i),\overline{A}(j))_{++})^*
\end{eqnarray}
The sets $(A(i),\overline{A}(j))_{-+}$ and $(A(i),\overline{A}(j))_{--}$ are complex conjugate of sets $(A(i),\overline{A}(j))_{+-}$ and $(A(i),\overline{A}(j))_{++}$, respectively. Only the measurements with planar target polarization can select uniquely one of the solutions for the phases~\cite{svec07b}.\\ 

For any given set of signs $\epsilon \overline{\epsilon}$ of the phases we can now calculate the corresponding helicity amplitudes $A_n(ij)_{\epsilon \overline{\epsilon}},i,j=1,2$. Using the expressions (2.5) and omitting the indices $\epsilon \overline{\epsilon}$ for brevity, the
helicity amplitudes read
\begin{eqnarray}
A_n(11) & = & {(-i)^n \over{\sqrt{2}}}(A(1)+(-1)^n \overline{A}(1)
\exp(i \omega_{11}))\exp(i\Phi_{S_u}(1))\\
\nonumber
A_n(12) & = & {(-i)^n \over{\sqrt{2}}}(A(1)+(-1)^n \overline{A}(2)
\exp(i \omega_{12}))\exp(i\Phi_{S_u}(1))\\
\nonumber
A_n(21) & = & {(-i)^n \over{\sqrt{2}}}(A(2)+(-1)^n \overline{A}(1)
\exp(i \omega_{21}))\exp(i\xi)\exp(i\Phi_{S_u}(1))\\
\nonumber
A_n(22) & = & {(-i)^n \over{\sqrt{2}}}(A(2)+(-1)^n \overline{A}(2)
\exp(i \omega_{22}))\exp(i\xi)\exp(i\Phi_{S_u}(1))
\end{eqnarray}
It is easy to verify that the effect of change of signs  of phases is again a complex conjugation of helicity amplitudes associated with a change of sign. Specifically, for any solution set $i,j$ we obtain
\begin{equation}
A_0(ij)_{-+}=+(A_0(ij)_{+-})^*, \qquad A_0(ij)_{--}=+(A_0(ij)_{++})^*
\end{equation}
\[
A_1(ij)_{-+}=-(A_1(ij)_{+-})^*, \qquad A_1(ij)_{--}=-(A_1(ij)_{++})^*
\]
The change of signs of $S$-wave phases results in change signs of relative phases $\omega$ and $\xi$
\begin{equation}
\omega_{ij,-+}=-\omega_{ij,+-}, \qquad \omega_{ij,--} = -\omega_{ij,++}
\end{equation}  
\[
\xi_{-+}=-\xi_{+-}, \qquad \xi_{--}=-\xi_{++}
\]
Since with the phase sets $--$ and $-+$ we do not get any new solutions, just complex conjugates of $++$ and $+-$ solutions, numerical calculations were done only for the sets with phases $++$ and $+-$.\\

Next we look at bilinear terms $A_nA_n^*=|A_n|^2$, $A=S,L,U$ and $A_nB_n^*$,$AB=LS,US,UL$.
For the sake of brevity we shall omit in the following the indices $ij$ and $++$, $+-$. Using (4.2) or (4.4) we obtain
\begin{eqnarray}
|A_n|^2 & = & {1\over{2}} \Bigl (|A|^2+|\overline{A}|^2+(-1)^n2X_A \cos(\omega)+
(-1)^n2Y_A \sin(\omega) \Bigr )\\
\nonumber
& = & {1\over{2}} \Bigl (I_A+(-1)^n2X_A \cos(\omega)+(-1)^n2Y_A \sin(\omega) \Bigr )
\end{eqnarray}
where $X_A=Re(A \overline{A}^*)$, $Y_A=Im(A \overline{A}^*)$ and $I_A=|A|^2+|\overline{A}|^2$
is partial wave intensity. Note that $Y_S=Im(S \overline{S}^*)=0$ as both $S$ and $\overline{S}$ are real. For the bilinear terms $A_nB_n^*$ we obtain
\begin{equation}
A_nB_n^*={1\over{2}} \Bigl ( AB^* + \overline{A}\overline{B}^* +(-1)^n \bigl (  A \overline{B}^* e^{-i\omega} +\overline{A}B^* e^{+i\omega} \bigr ) \Bigr )
\end{equation}
The real part reads
\begin{equation}
Re(A_n B_n^*)={1\over{2}} \Bigl ( Re(AB^*)+Re(\overline{A} \overline{B}^*)
\end{equation}
\[
+(-1)^n \bigl ( (Re(A\overline{B}^*)+Re(\overline{A}B^*) ) \cos\omega
+(Im(A \overline{B}^*)-Im(\overline{A} B^*) )\sin \omega \bigr ) \Bigr )
\]
The imaginary part reads
\begin{equation}
Im(A_n B_n^*)={1\over{2}} \Bigl ( Im(AB^*)+Im(\overline{A} \overline{B}^*)
\end{equation}
\[
+(-1)^n \bigl ( (Im(A\overline{B}^*)+Im(\overline{A}B^*) ) \cos\omega
-(Re(A \overline{B}^*)-Re(\overline{A} B^*) )\sin \omega \bigr ) \Bigr )
\]

It is apparent from (4.2) and (4.4) that the knowledge of $\omega$ allows to determine the helicity amplitudes up to an absolute phase for each solution set $i,j$ and phase set $++$ and $+-$. As we show in the next Section, the phase $\omega$ can be determined analytically from the consistency condition
\begin{equation}
|A_n|^2|B_n|^2=(Re(AB^*))^2+(Im(AB^*))^2
\end{equation}
where the terms are given by (4.7),(4.9) and (4.10).

\section{Analytical solutions of the relative phase $\omega$.}

In order to make use of the consistency condition (4.11) to determine $\omega$ , we first rewrite (4.7), (4.9) and (4.10) in a simplified form. To this end we define
\begin{eqnarray}
X(AB) & = & Re(AB^*)+Re(\overline{A} \overline{B}^*)\\
\nonumber
Y(AB) & = & Im(AB^*)+Im(\overline{A} \overline{B}^*)\\
\nonumber
X(A\overline{B})_+ & = & Re(A\overline{B}^*)+Re(\overline{A}B^*)\\
\nonumber
Y(A\overline{B})_+ & = & Im(A\overline{B}^*)+Im(\overline{A}B^*)\\
\nonumber
X(A\overline{B})_- & = & Re(A\overline{B}^*)-Re(\overline{A}B^*)\\
\nonumber
Y(A\overline{B})_- & = & Im(A\overline{B}^*)-Im(\overline{A}B^*)
\end{eqnarray}
Then
\begin{eqnarray}
Re(A_nB_n^*) & = & {1\over{2}} \Bigl ( X(AB)+(-1)^n \bigl (X(A\overline{B})_+ \cos \omega +Y(A\overline{B})_- \sin \omega \bigr ) \Bigr )\\
\nonumber
Im(A_nB_n^*) & = & {1\over{2}} \Bigl ( Y(AB)+(-1)^n \bigl (Y(A\overline{B})_+ \cos \omega -X(A\overline{B})_- \sin \omega \bigr ) \Bigr )\\
\nonumber
Re(A_nA_n^*)=|A_n|^2 & = & {1\over{2}} \Bigl ( X(AA)+(-1)^n \bigl (X(A\overline{A})_+ \cos \omega +Y(A\overline{A})_- \sin \omega \bigr ) \Bigr )
\end{eqnarray}
Note that $X(AA)=I_A$, $X(A\overline{A})_+=2X_A$ and $Y(A\overline{A})_-=2Y_A$. We can write
\begin{eqnarray}
Re(A_nB_n^*) & = & {1\over{2}} \Bigl ( X(AB)+(-1)^n XG(A \overline{B}) \Bigr )\\
\nonumber
Im(A_nB_n^*) & = & {1\over{2}} \Bigl ( Y(AB)+(-1)^n YG(A \overline{B}) \Bigr )\\
\nonumber
     |A_n|^2 & = & {1\over{2}} \Bigl ( X(AA)+(-1)^n XG(A \overline{A}) \Bigr )\\
\nonumber
     |B_n|^2 & = & {1\over{2}} \Bigl ( X(BB)+(-1)^n XG(B \overline{B}) \Bigr )
\end{eqnarray}
where
\begin{eqnarray}
XG(A \overline{B}) & = & X(A\overline{B})_+ \cos \omega +Y(A\overline{B})_- \sin \omega\\
\nonumber
YG(A \overline{B}) & = & Y(A\overline{B})_+ \cos \omega -X(A\overline{B})_- \sin \omega 
\end{eqnarray}
Next we require that
\begin{equation}
|A_n|^2|B_n|^2=(Re(AB^*))^2+(Im(AB^*))^2
\end{equation}
The l.h.s. of (5.5) reads
\begin{equation}
X(AA)X(BB)+XG(A\overline{A})XG(B\overline{B}) 
\end{equation}
\[
+(-1)^n X(AA)XG(A\overline{A})X+(-1)^nX(BB)XG(A\overline{A})
\]
The r.h.s. of (5.5) reads 
\begin{equation}
X(AB)^2+XG(A \overline{B})^2+Y(AB)^2+YG(A \overline{B})^2
\end{equation}
\[
+(-1)^n2X(AB)XG(A \overline{B})+(-1)^n2Y(AB)YG(A \overline{B})
\]
Subtracting (5.5) with $n=1$ from (5.5) with $n=0$ and using (5.4) we obtain equation linear in $\cos \omega$ and $\sin \omega$
\begin{equation}
\sin \omega \Bigl ( X(AA)Y(B\overline{B})_- +X(BB)Y(A\overline{A})_- -2X(AB)Y(A\overline{B})_- +2Y(AB)X(A\overline{B})_- \Bigr ) =
\end{equation}
\[
-\cos \omega \Bigl ( X(AA)X(B\overline{B})_+ +X(BB)X(A\overline{A})_+ -2X(AB)X(A\overline{B})_+ -2Y(AB)Y(A\overline{B})_+ \Bigr )
\]
which can be cast in the form
\begin{equation}
\sin \omega W_2 = - \cos \omega W_1
\end{equation}
Using $\sin^2 \omega + \cos^2 \omega =1$ we find
\begin{eqnarray}
\cos \omega & = & {\pm W_2 \over{W}}\\
\nonumber
\sin \omega & = & {\mp W_1 \over{W}}
\end{eqnarray}
where $W=\sqrt{W_1^2+W_2^2}$.\\

Using (4.7), (4.9) and (4.10) it is straightforward to verify that the consistency condition (5.5) reduces to identity in the following two cases
\begin{eqnarray}
\cos \omega =0, & \sin \omega = \pm 1\\
\sin \omega =0, & \cos \omega = \pm 1
\end{eqnarray}
The first case leads to two solutions $\omega = + \pi/2$ and $\omega= -\pi/2$. The second case leads to another two solutions $\omega=0$ and $\omega=\pi$.

\section{Solutions for $S$- and $P$-wave helicity amplitudes.}

\subsection{Numerical calculations and their checks.}

Monte Carlo amplitude analysis of CERN data on $\pi^- p \to \pi^- \pi^+ n$ measured on transversely polarized target at 17.2 GeV/c at small momentum transfers $0.005 \leq -t \leq 0.2$ (GeV/c)$^2$ was performed using a computer code AACERNM for dipion masses 580 -1080 MeV~\cite{svec07b}. The mean values of moduli and phases of reduced transversity amplitudes were used as an input in a computer code HACERNMa to calculate $\omega$ and helicity amplitudes $A_n,n=0,1$, $A=S,L,U$. The mean values of moduli and phases of the reduced transversity amplitudes satisfy strict normalization and phase conditions, respectively, and thus represent the true measured amplitudes at the $t$-bin average of 0.067 (GeV/c)$^2$~\cite{svec02a,svec07b}.\\

Equations (5.10) were used with amplitudes $A=L,B=S$ to calculate $\omega(ij)$ for each set $++$ and $+-$ of signs of phases. Moduli $|A_n|^2,|B_n|^2$ and interference terms $Re(A_nB_n^*),Im(A_nB_n^*)$ for nonflip $n=0$ and flip $n=1$ pairs of amplitudes $AB=LS,US,UL$ were calculated to determine cosines and sines of the relative phases 
\begin{equation}
\Phi(A_nB_n^*)=\Phi_{A_n}-\Phi_{B_n}
\end{equation}
Since the calculations of $\omega$, moduli and the interference terms are all entirely independent, the selfconsistency of the data and calculations was checked using the following three tests on the relative phases. The first test are trigonometric identities
for $n=0,1$ for each pair of amplitudes $AB=LS,US,UL$
\begin{equation}
\cos^2 \Phi(A_nB_n^*) + \sin^2 \Phi(A_nB_n^*)=1
\end{equation}
The second test are phase conditions for $n=0,1$
\begin{equation}
(\Phi_{L_n}-\Phi_{S_n})-(\Phi_{U_n}-\Phi_{S_n})+(\Phi_{U_n}-\Phi_{L_n})=0
\end{equation}
The third test are cosine conditions for $n=0,1$ equivalent to phase conditions
\begin{equation}
\cos^2 \Phi(L_nS_n^*)+\cos^2 \Phi(U_nS_n^*)+\cos^2 \Phi(U_nL_n^*)-
2\cos \Phi(L_nS_n^*)\cos \Phi(U_nS_n^*)\cos \Phi(U_nL_n^*)=1
\end{equation}
All conditions are satisfied identically within the single precision calculation used by HACERNMa for all combinations of solutions $i,j=1,2$ for both sets $++$ and $+-$ of signs of phases. For instance, typical deviation from 1 of the trigonometric identities (6.2) are of order $10^{-5} - 10^{-7}$.\\

Errors on $\omega$ and helicity amplitudes were not calculated. This could be accomplished by combining the core of AACERNM with the code HACERNMa and calculating the helicity amplitudes for each Monte Carlo sampling of the data that yields physical solutions for the reduced transversity amplitudes. The resulting distributions of values of helicity amplitudes would define their mean values, and the range of distributions would estimate their errors. These calculations are however beyond the scope and purpose of this paper.

\subsection{Unphysical solutions with $\cos \omega \neq 0$ and $\sin \omega \neq 0$.}

The calculated values of $\omega(ij)_{++}$ and $\omega(ij)_{+-}$ show random variations as a function of dipion mass for all solution sets $i,j=1,2$ and signs of phases. To test the consistency condition (2.7)
\begin{equation}
\xi=\omega_{11}-\omega_{21}=\omega_{12}-\omega_{22}
\end{equation}
we calculated the quantity
\begin{equation}
\Delta=(\omega_{11}-\omega_{21})-(\omega_{12}-\omega_{22})
\end{equation}
The results are shown in Fig. 1. The large random variations of $\Delta$ indicate a strong violation of the consistency condition which requires $\Delta \equiv 0$. These results are thus inconsistent with Kraus representation and the assumption that all four sets of reduced transversity amplitudes are physical.\\ 

One could hope that one of the solutions could be selected as a valid physical solution. However, all amplitudes exhibit the same random behaviour and do not show the required resonant Breit-Wigner behaviour at $\rho^0(770)$ resonance. In Fig. 2 and Fig. 3 we present the results for the single flip amplitude $L_1$ for the signs of phases $++$ and $+-$, respectively, with $\cos\omega=-W_2/W$ to illustrate this behaviour. Moreover, the non-flip amplitude $L_0$ has magnitude comparable to or larger than the flip amplitude $L_1$, in contradiction with the physical requirement that $L_1$ dominates on account of pion exchange dominance at small $t$. The results with $\cos\omega=+W_2/W$ are similar. To ensure the positivity of $Im L_1$ the absolute phases $\Phi_{S_u}(i)$ were set equal to $-\pi$.\\

On the basis of the unphysical behaviour of the helicity amplitudes, all solutions with $\cos \omega = \mp W_2/W$ are considered unphysical. They are excluded not only as co-evolution amplitudes but also as possible standard amplitudes.

\subsection{Unphysical solutions with $\cos \omega = 0$ and $\sin \omega = \pm 1$.}

The two values for $\sin \omega = \pm 1$ correspond to $\omega_{ij,++}=\pm \pi/2$ and $\omega_{ij,+-}= \pm \pi/2$ for all solution sets $i,j=1,2$ and signs of phases. Both solutions satisly the consistency condition (2.7).\\

At small $t$ the pion creation process $\pi^- p \to \pi^- \pi^+ n$ is dominated by pion exchange in the $t$ channel. The pion exchange amplidues are the unnatural exchange helicity flip amplitudes $S_1,L_1$ and $U_1$ which must dominate the helicity non-flip amplitudes $S_0,L_0$ and $U_0$, respectively. The CERN data correspond to small $t$ bin average $-t$=0.067 (GeV/c)$^2$. As the result, pion exchange dominance of the flip helicity amplitudes must be observed by the solutions.\\

The solutions with $\sin \omega =+1$ are excluded because they require that the non-flip amplitude $L_0$ is larger than the flip amplitude $L_1$. Because $\cos \omega=0$ and the amplitudes $S_0$ and $S_1$ do not depend on $\sin \omega$, both solutions $\sin \omega =\pm 1$ give $|S_0|^2=|S_1|^2=I_S/2$ where $I_S$ is $S$-wave intensity. The equal magnitudes of the $S$-wave helicity flip and non-flip amplitudes contradict the pion exchange dominance for both solutions for $\omega$. As a result, both solutions are unphysical for any solution set $i,j=1,2$ and signs of phases $++$ and $+-$ and are excluded from consideration as co-evolution or standard amplitudes.\\

For the sake of comparison we present some results for $\sin \omega=-1$. Fig. 4 shows the amplitudes $S_1$ which are the same for both signs of phases $++$ and $+-$. In Fig. 5 and Fig. 6 we show the amplitudes  $L_1$ for the signs of phases $++$ and $+-$, respectively.
To ensure the positivity of $Im L_1$ the absolute phases $\Phi_{S_u}(i)$ were set equal to $+\pi/2$ ($-\pi/2$ for $\sin \omega =+1$). The imaginary parts of $L_1$ exhibit a clear $\rho^0(770)$ resonant structure. There is no evidence of a chaotic behaviour in any of the amplitudes.

\subsection{Physical solutions with $\cos \omega = -1$ and $\sin \omega = 0$.}

The two values for $\cos \omega = \pm 1$ correspond to $\omega_{ij,++}=0,\pi$ and $\omega_{ij,+-}= 0,\pi$ for all solution sets $i,j=1,2$ and signs of phases. Both solutions satisly the consistency condition (2.7) with $\xi \equiv 0$. From the definition (2.6) of $\xi=\Phi_{S_u}(2)-\Phi_{S_u}(1)$ it follows that the amplitudes $S_u(1)$ and $S_u(2)$ must have the same phase.\\

The solutions with $\cos \omega=+1$ ($\omega=0$) require that $|S_1|^2 <|S_0|^2$ and $|L_1|^2<|L_0|^2$ for any solution set $i,j=1,2$ and signs of phases. The magnitudes of $S_1$ are small compared to magnitudes of $S_0$, in contradiction with the pion exchange dominance. These solutions are thus excluded as unphysical solutions.\\

In the solutions with $\cos \omega =-1$ the non-flip and flip amplitudes are interchenged and the pion exchange dominance is observed. To ensure the positivity of $Im L_1$ the absolute phases $\Phi_{S_u}(i)$ were set equal to $-\pi$. The resulting solutions represent a unique solution for both helicity and transversity amplitudes (up to signs of phases to be resolved by measurements with planar components of target polarization). The amplitudes are consistent with Kraus representation and with the the central hypothesis (1.1) and can thus be identified with a unique set of co-evolution amplitudes.\\ 

The results for helicity flip amplitudes are presented in Figures 7-11. The Fig. 7 shows the amplitude $S_1$ for both sets of signs $++$ and $+-$ of phases. While there is an indication for a $\rho^0(770)$ structure in solutions $S_1(11)$ and $S_1(21)$, the solutions $S_1(12)$ and $S_1(22)$ show a broad structure indicating apparent suppression of $\rho^0(770)$. All solutions for $S_1$ show expected resonant structures corresponding to $f_0(980)$ resonance. The Figures 8 and 9 show the amplitude $L_1$ for signs of phases $++$ and $+-$, respectively, while the Figures 10 and 11 show the same for amplitudes $U_1$.  Both $Im L_1$ and $Im U_1$ show clear resonant peaks at $\rho^0(770)$ mass but have opposite signs. Their real parts are relatively structureless and small. Important structures appear in $Im L_1 $ at $f_0(980)$ mass consistent with $\rho^0(770)-f_0(980)$ mixing~\cite{svec07b}.\\

The results for helicity non-flip amplitudes are presented in Figures 12-16. The Fig. 12 shows the amplitude $S_0$ for both sets of signs $++$ and $+-$ of phases. The non-flip amplitude $S_0$ is structureless and very small compared to flip amplitude $S_1$. The Figures 13 and 14 show the amplitude $L_0$ for signs of phases $++$ and $+-$, respectively, while the Figures 15 and 16 show the same for amplitudes $U_0$. Solutions for both amplitudes appear structureless at $\rho^0(775)$ mass range with $Im L_0$ changing sign around $f_0(980)$ mass. The non-flip amplitudes $U_0$ are very small compared to flip amplitudes $U_1$.\\

The solutions with $\cos \omega =-1$ represent the first model independent determination of all unnatural exchange helicity amplitudes in $\pi^- p \to \pi^- \pi^+ n$. Assuming the resolution of sign ambiguity of phases in measurements with planar components of target polarization, the solution is unique without ambiguity. Moreover, assuming continuity of $\omega$ and its constant value of $\pi$ at larger momentum transfers $t$ or dipion masses $m$, this solution can be selected in all amplitude analyses to provide a unique solution of co-evolution amplitudes at any $t$ and $m$.

\subsection{The phases of amplitudes $L_1$ and $S_1$.}

In the physical solutions with $\cos \omega=-1$ the phase of amplitude $S_1$ is $\Phi_{S_1}(ij)=90^{\circ}$. The phase of amplitude $L_1$ is shown in Fig. 17. It is nearly constant at $\sim 90^{\circ}$ below 950 MeV, which appears suprising as one would expect its phase near the phase of $\rho^0(770)$ resonamce. Its sudden change of sign above 950 MeV is due to sudden change of sign in density matrix elements $(\rho^0_u)_{0s}$, $(\rho^0_u)_{1s}$, $(\rho^0_y)_{0s}$, and  $(\rho^y_u)_{1s}$ above this mass.\\

We can assign to the amplitude $L_1$ the Breit-Wigner phase $\Phi(\rho^0)$ of $\rho^0(770)$ resonance by modifying the absolute phase $-\pi \to -\pi+\Phi(\rho^0)-\Phi_{L_1}(ij)$. This uniformly modifies the phases of all amplitudes. In particular, the amplitudes $S_1$ acquire a modified phase
\begin{equation}
\Phi_{S_1}(ij)'=\Phi_{S_1}(ij) + \Phi(\rho^0)-\Phi_{L_1}(ij)
\end{equation}
Since $\Phi_{S_1}(ij) - \Phi_{L_1}(ij) \approx 0$, the modified phase of $S_1$ becomes a somewhat modified phase of $\Phi(\rho^0)$.  The modified phase of $S_1$ is shown in Fig. 18. The modification depends on the sign of the phases $++$ and $+-$.\\

For the sake of comparison we show in Fig. 19 the phase of $L_1$ for the solutions $\sin \omega=-1$.  The phases are again approximately constant below 950 MeV and can be assigned the Breit-Wigner phase of $\rho^0(770)$. The modified phases of amplitudes $S_1$ are again near the phase $\Phi(\rho^0)$, as shown in Fig. 20.\\

In 1997 Kaminski, Lesniak and Rybicki published a model dependent determination of helicity amplitudes based on CERN-Munich-Cracow analysis of the CERN data which did not determine the phases of the reduced transversity amplitudes but only their cosines~\cite{kaminski97}.
They assumed Breit-Wigner phase for the amplitudes $L$ and $\overline{L}$ and that the phases $\Phi_{LS}$ and $\overline{\Phi}_{LS}$ change sign near $\rho^0(770)$ mass. For the phases of amplitudes $S_1$ they obtained two "steep" solution and two "flat" solutions. Their "steep" solutions are near $\Phi(\rho^0)$ phase and are similar to our results in Fig. 18. Their "flat" solutions have much smaller rise and correspond closely to the CERN-Munich $\pi \pi$ phases shifts $\delta^0_0$. The authors emphasize in their paper that they can get the "flat" solutions only provided that the phases $\Phi_{LS}$ and $\overline{\Phi}_{LS}$ change sign near $\rho^0(770)$ mass. In our high definition amplitude analysis~\cite{svec07b} we have determined the phases $\Phi_{LS}$ and $\overline{\Phi}_{LS}$ and find no evidence for such a change of sign. Moreover, we show that all eigenvalues of the reduced density matrix are non-zero except at the point where the phase $\Phi_{LS}$ or $\overline{\Phi}_{LS}$ change sign and where some of the eigenvalues vanish. Furthermore, the change of sign of the phases $\Phi_{LS}$ and $\overline{\Phi}_{LS}$ causes abrupt and discontinous change of sign in all other phases~\cite{svec07b}. It is the unchangig sign of the phases $\Phi_{LS}$ and $\overline{\Phi}_{LS}$ which prevents such anomalous behaviours of the eigenvalues and other phases which are not considered in Ref.~\cite{kaminski97}. We suggest that their analysis without the assumption of the change of sign of these phases would be similar to our model independent results.

\section{Uniqueness of Kraus representation of the mixed final state density matrix.}

The unique solution for the phase $\omega = \pi$ implies a unique solution for the $S$- and $P$-wave transversity and helicity amplitudes from measurements on polarized target. In this Section we show that the four solutions for the $S$- and $P$-wave final state density matrices $\rho_f(ij),i,j=1,2$ are uniquely determined and, as a result, so is the Kraus representation of the mixed final state density matrix
\begin{equation}
\rho_f=p_{11}\rho_f(11)+p_{12}\rho_f(12)+p_{21}\rho_f(21)+p_{22}\rho_f(22)
\end{equation} 
Extension to higher dimeson masses where $D$-waves contribute will be given elsewhere.\\

For the specific case of $\pi^- p \to \pi^- \pi^+ n$ and similar processes with target polarization $\vec{P}=(P_x,P_y,P_z)$ the Kraus representation (3.3) takes the more explicit form
\begin{equation}
\rho_f(\theta \phi,\vec{P})=p_{11}\rho_f(11,\theta \phi,\vec{P})+p_{12}\rho_f(12,\theta \phi,\vec{P})+p_{21}\rho_f(21,\theta \phi,\vec{P})+p_{22}\rho_f(22,\theta \phi,\vec{P})
\end{equation}
where
\begin{equation}
\rho_f(ij,\theta \phi,\vec{P})=T(ij,\theta \phi) \rho_p(\vec{P})T(ij.\theta \phi)^+
\end{equation}
In (7.3) $T(ij,\theta \phi)$ is the matrix of co-evolution transition amplitudes defined by (3.7)
\begin{equation}
<\theta \phi,\chi|<ij|T|ij>|0,\nu>
\end{equation}
where $\Omega=\theta \phi$ describes the direction of $\pi^-$ in the dipion center-of-mass system, $\chi$ is the recoil nucleon helicity, $\nu$ is the target nuleon helicity and $0$ stands for incident pion helicity. The momenta of the particles are suppressed. In (7.3) $\rho_p(\vec{P})$ is the density matrix of target nucleons
\begin{equation}
\rho_p(\vec{P})={1\over{2}}(\sigma_u+\vec{P} \vec{\sigma})
\end{equation}
where $\vec{\sigma}$ are Pauli matrices and $\sigma_u=1$ is a unit matrix.\\

Each final state density matrix $\rho_f(ij,\theta \phi, \vec{P})$ is a single qubit density matrix corresponding to spin ${1 \over{2}}$ of the recoil nucleon. It can be written in the form~\cite{nielsen00,svec07a}
\begin{equation}
\rho_f(ij,\theta \phi, \vec{P}) = {1 \over{2}} (I^0 (ij,\theta \phi, \vec{P}) \sigma^0 + {\vec{I}} (ij,\theta \phi, \vec{P}) \vec{\sigma})
\end{equation}
\noindent
where the traces $I^\ell(ij,\theta \phi, \vec{P}) = Tr(\sigma^\ell \rho_f(ij,\theta \phi, \vec{P})), \ell=0,1,2,3$ represent measurable intensities of angular distributions. Using $\rho_p={1 \over {2}}(1+\vec{P}\vec{\sigma})$ we can write matrix elements of $\rho_f(ij,\theta \phi, \vec{P})$ in terms of components of target polarization
\begin{equation}
(\rho_f(ij,\theta \phi, \vec{P}))^{{1 \over{2}}{1 \over{2}}}_{\chi \chi'}= (\rho_u(ij,\theta\phi))^{{1 \over{2}}{1 \over{2}}}_{\chi\chi^{'}}+P_x(\rho_x(ij,\theta \phi))^{{1 \over{2}}{1 \over{2}}}_{\chi\chi^{'}}+P_y(\rho_y(ij,\theta \phi))^{{1 \over{2}}{1 \over{2}}}_{\chi \chi^{'}}+P_z(\rho_z(ij,\theta \phi))^{{1 \over{2}}{1 \over{2}}}_{\chi \chi^{'}}
\end{equation}
In (7.7) the subscript $u$ stands for unpolarized target $\vec{P} = 0$. Using the decomposition (7.7) of $\rho_f(ij,\theta \phi, \vec{P})$  we find a decomposition for the intensities
\begin{equation}
I^j(ij,\theta \phi, \vec{P}) = Tr(\sigma^\ell \rho_f(ij,\theta \phi, \vec{P}))=I^\ell_u(ij,\theta \phi) + P_x I^\ell_x(ij,\theta \phi)+ P_y I^\ell_y(ij,\theta \phi)+ P_z I^\ell_z(ij,\theta \phi)
\end{equation}
where the components $I^\ell_k(ij,\theta \phi)$, $\ell=0,1,2,3$ and $k=u,x,y,z$ of the intensities $I^\ell(\theta \phi, \vec{P})$ are given by traces
\begin{equation}
I^\ell_k(ij,\theta \phi) = Tr_{\chi, \chi''} ((\sigma^\ell)_{\chi \chi''} (\rho_k(ij,\theta \phi))^{{1 \over{2}}{1 \over{2}}}_{\chi'' \chi})
\end{equation}

\begin{table}
\caption{Density matrix elements $Re \rho^0_u {{d^2 \sigma}/{dtdm}}$, $Re \rho^0_y {{d^2 \sigma}/{dtdm}}$ and $Re \rho^2_u {{d^2 \sigma}/{dtdm}}$, $Re \rho^2_y {{d^2 \sigma}/{dtdm}}$ in terms of reduced transversity amplitudes. Here $\sigma_A=|A|^2+|\bar {A}|^2$ and $\tau_A=|A|^2-|\bar {A}|^2$ for $A=U,N$. The expressions are valid for any  $\omega$.}
\begin{tabular}{ccccc}
\toprule
$\rho^{JJ'}_{\lambda \lambda}$&$Re \rho^0_u$&$Re \rho^0_y$&$Re \rho^2_u$&$Re \rho^2_y$\\
\colrule
$\rho^{00}_{ss}$&$|S|^2+|\bar {S}|^2$&$|S|^2-|\bar {S}|^2$ &-$(\rho^0_y)^{00}_{ss}$&-$(\rho^0_u)^{00}_{ss}$\\
$\rho^{11}_{00}$&$|L|^2+|\bar {L}|^2$&$|L|^2-|\bar {L}|^2$
&-$(\rho^0_y)^{11}_{00}$&-$(\rho^0_u)^{11}_{00}$\\
$\rho^{11}_{11}$&${1\over{2}}(\sigma_U+\sigma_N)$&${1\over{2}}(\tau_U+\tau_N)$
&-($\rho^0_y)^{11}_{11}+\tau_N$&-($\rho^0_u)^{11}_{11}+\sigma_N$\\
$\rho^{11}_{1-1}$&$-{1\over{2}}(\sigma_U-\sigma_N)$&$-{1\over{2}}(\tau_U-\tau_N)$
&-$(\rho^0_y)^{11}_{1-1}+\tau_N$&-$(\rho^0_u)^{11}_{1-1}+\sigma_N$\\
$Re \rho^{10}_{0s}$&$Re (L S^*+\bar {L} \bar {S}^*)$&$Re (LS^*-\bar {L} \bar {S}^*)$
&-$Re (\rho^0_y)^{10}_{0s}$&-$Re (\rho^0_u)^{10}_{0s}$\\
$\sqrt{2}Re \rho^{10}_{1s}$&$Re (US^*+\bar {U} \bar {S}^*)$&$Re (US^*-\bar {U} \bar {S}^*)$&-$\sqrt{2}Re (\rho^0_y)^{10}_{1s}$&-$\sqrt{2}Re (\rho^0_u)^{10}_{1s}$\\
$\sqrt{2}Re \rho^{11}_{01}$&$Re (LU^*+\bar {L} \bar {U}^*)$&$Re (LU^*-\bar {L} \bar {U}^*)$&-$\sqrt{2}Re (\rho^0_y)^{11}_{01}$&-$\sqrt{2}Re (\rho^0_u)^{11}_{01}$\\
\botrule
\end{tabular}
\label{Table I.}
\end{table}

\begin{table}
\caption{Density matrix elements $Im \rho^0_x {{d^2 \sigma}/{dtdm}}$, $Im \rho^0_z {{d^2 \sigma}/{dtdm}}$ and $Im \rho^2_x {{d^2 \sigma}/{dtdm}}$, $Im \rho^2_z {{d^2 \sigma}/{dtdm}}$ in terms of reduced transversity amplitudes. The expressions are valid for any $\omega$.}
\begin{tabular}{ccccc}
\toprule
$\rho^{JJ'}_{\lambda \lambda}$&$Im \rho^0_x$&$Im \rho^0_z$&$Im \rho^2_x$&$Im \rho^2_z$\\
\colrule
$\sqrt{2}Im \rho^{01}_{s1}$&$Re(-S \bar {N}^*+N \bar {S}^*)$&$Im(+S \bar {N}^*-N \bar {S}^*)$&$Re(S \bar {N}^*+N \bar {S}^*)$&$Im(-S \bar {N}^*-N \bar {S}^*)$\\
$\sqrt{2}Im \rho^{11}_{01}$&$Re(-L \bar {N}^*+N \bar {L}^*)$&$Im(+L \bar {N}^*-N \bar {L}^*)$&$Re(L \bar {N}^*+N \bar {L}^*)$&$Im(-L \bar {N}^*-N \bar {L}^*)$\\
$Im \rho^{11}_{-11}$&$Re(+U \bar {N}^*-N \bar {U}^*)$&$Im(-U \bar {N}^*+N \bar {S}^*)$
&$Re(-U \bar {N}^*-N \bar {U}^*)$&$Im(+U \bar {N}^*+N \bar {S}^*)$\\
\botrule
\end{tabular}
\label{Table II.}
\end{table}

\begin{table}
\caption{Density matrix elements $Re \rho^1_x {{d^2 \sigma}/{dtdm}}$, $Re \rho^1_z {{d^2 \sigma}/{dtdm}}$ and $Re \rho^3_x {{d^2 \sigma}/{dtdm}}$, $Re \rho^3_z {{d^2 \sigma}/{dtdm}}$ in terms of reduced transversity amplitudes. The expressions are valid for $\omega=\pi$.}
\begin{tabular}{ccccc}
\toprule
$\rho^{JJ'}_{\lambda \lambda}$&$Re \rho^1_x$&$Re \rho^1_z$&$Re \rho^3_x$&$Re \rho^3_z$\\
\colrule
$\rho^{00}_{ss}$&$2Re(S \overline{S}^*)$&-$2Im(S \overline{S}^*)$
&+$(\rho^1_z)^{00}_{ss}$&-$(\rho^1_x)^{00}_{ss}$\\
$\rho^{11}_{00}$&$2Re(L \overline{L}^*)$&-$2Im(L \overline{L}^*)$
&+$(\rho^1_z)^{11}_{00}$&-$(\rho^1_x)^{11}_{00}$\\
$\rho^{11}_{11}$&$Re(+U \overline{U}^*-N\overline{N}^*)$&
$Im(-U \overline{U}^*+N\overline{N}^*)$
&+($\rho^1_z)^{11}_{11}-2Im(N\overline{N}^*)$&-($\rho^1_x)^{11}_{11}-2Re(N\overline{N}^*)$\\
$\rho^{11}_{1-1}$&$Re(-U \overline{U}^*-N\overline{N}^*)$&
$Im(+U \overline{U}^*+N\overline{N}^*)$
&+($\rho^1_z)^{11}_{1-1}-2Im(N\overline{N}^*)$&-($\rho^1_x)^{11}_{1-1}-2Re(N\overline{N}^*)$\\
$Re \rho^{10}_{0s}$&$Re(L\overline{S}^*+S\overline{L}^*)$&
$Im(-L\overline{S}^*+S\overline{L}^*)$
&+$Re (\rho^1_z)^{10}_{0s}$&-$Re (\rho^1_x)^{10}_{0s}$\\
$\sqrt{2}Re \rho^{10}_{1s}$&$Re(U\overline{S}^*+S\overline{U}^*)$&
$Im(-U\overline{S}^*+S\overline{U}^*)$
&+$\sqrt{2}Re (\rho^1_z)^{10}_{1s}$&-$\sqrt{2}Re (\rho^1_x)^{10}_{1s}$\\
$\sqrt{2}Re \rho^{11}_{10}$&$Re(U\overline{L}^*+L\overline{U}^*)$&
$Im(-U\overline{L}^*+L\overline{U}^*)$
&+$\sqrt{2}Re (\rho^1_z)^{11}_{10}$&-$\sqrt{2}Re (\rho^1_x)^{11}_{10}$\\
\botrule
\end{tabular}
\label{Table III.}
\end{table}

\begin{table}
\caption{Density matrix elements $Im \rho^1_u {{d^2 \sigma}/{dtdm}}$, $Im \rho^1_y {{d^2 \sigma}/{dtdm}}$ and $Im \rho^3_u {{d^2 \sigma}/{dtdm}}$, $Im \rho^3_y {{d^2 \sigma}/{dtdm}}$ in terms of reduced transversity amplitudes. The expression are valid for $\omega = \pi$.}
\begin{tabular}{ccccc}
\toprule
$\rho^{JJ'}_{\lambda \lambda}$&$Im \rho^1_u$&$Im \rho^1_y$&$Im \rho^3_u$&$Im \rho^3_y$\\
\colrule
$\sqrt{2}Im \rho^{01}_{s1}$
&$Re(SN^*-\overline{S}\overline{N}^*)$&$Re(SN^*+\overline{S}\overline{N}^*)$
&$Im(-SN^*-\overline{S}\overline{N}^*)$&$Im (-SN^*+\overline{S}\overline{N}^*)$\\
$\sqrt{2}Im \rho^{11}_{01}$
&$Re(LN^*-\overline{L}\overline{N}^*)$&$Re(LN^*+\overline{L}\overline{N}^*)$
&$Im(-LN^*-\overline{L}\overline{N}^*)$&$Im (-LN^*+\overline{L}\overline{N}^*)$\\
$Im \rho^{11}_{-11}$
&$Re(-UN^*+\overline{U}\overline{N}^*)$&$Re(-UN^8-\overline{U}\overline{N}^*)$
&$Im(+UN^*+\overline{U}\overline{N}^*)$&$Im (+UN^8-\overline{U}\overline{N}^*)$\\
\botrule
\end{tabular}
\label{Table IV.}
\end{table}

In general, a plane wave helicity state of two particles with helicities $\mu_1, \mu_2$ can be expanded in terms of angular helicity states ~\cite{perl74,martin70}
\begin{equation}
|p\theta \phi;\mu_1 \mu_2>=\sum \limits_{J,\lambda} \sqrt{{2J+1} \over {4\pi}} D^J_{\lambda,\mu}(\phi,\theta,-\phi)|pJ\lambda;\mu_1 \mu_2>
\end{equation}
where p is the momentum in center-of-mass system and $J$ and $\lambda$ are the two-particle spin and helicity, and $\mu=\mu_1-\mu_2$. For two pions $\mu_1=\mu_2=0$ and $D^J_{\lambda0}(\phi,\theta,-\phi)=\sqrt{4\pi/(2J+1)}Y^{J*}_\lambda (\theta,\phi)$. The final state can be expanded in spherical harmonics
\begin{equation}
|\theta \phi, \chi> = \sum \limits_{J \lambda} Y^{J*}_{\lambda}(\theta, \phi) |J \lambda, \chi>
\end{equation}
where $J$ and $\lambda$ are dipion spin and helicity, respectively. Using (7.11) we obtain the angular expansion of transition amplitudes (7.4). The final expressions for $I^\ell_k(ij,\theta \phi)$ with independent angular density matrix elements assume $P$ parity conservation and read~\cite{svec07a}
\begin{equation}
I^\ell_k(ij,\theta \phi) = {{d^2 \sigma} \over{dtdm}}
\sum \limits_{J \leq J'} \sum \limits _{\lambda \geq 0} \sum \limits_{\lambda'} 
\xi_{JJ'} \xi_\lambda (Re \rho^\ell_k(ij))^{JJ}_{\lambda \lambda'} Re(Y^J_\lambda(\theta\phi)Y^{J*}_{\lambda'}(\theta \phi))
\end{equation}
for $(k,\ell)=(u,0),(y,0),(u,2),(y,2),(x,1),(z,1),(x,3),(z,3)$ and
\begin{equation}
I^\ell_k(ij,\theta \phi) = {{d^2 \sigma} \over{dtdm}} 
\sum \limits_{ J \leq J'} \sum \limits_{\lambda \geq 0} \sum \limits_{\lambda'} 
\xi_{JJ'} \xi_\lambda (Im \rho^\ell_k(ij))^{JJ'}_{\lambda \lambda'} 
Im(Y^J_\lambda(\theta \phi)Y^{J'*}_{\lambda'}(\theta \phi))
\end{equation}
for $(k,\ell)=(x,0),(z,0),(x,2),(z,2),(u,1),(y,1),(u,3),(y,3)$. In (7.12) and (7.13)  $\xi_0=1$ and $\xi_\lambda=2$ for $\lambda >0$. The factor $\xi_{JJ'}=1$ for $J=J'$ and $\xi_{JJ'}=2$ for $J<J'$.\\

Lutz and Rybicki tabulated expressions for density matrix elements $(R^\ell_k)^{JJ'}_{\lambda \lambda'}={{d^2 \sigma} \over{dtdm}}(\rho^\ell_k)^{JJ}_{\lambda \lambda'}$ in terms of helicity amplitudes, helicity amplitudes with definite $t$-channel naturality, transversity amplitudes and transversity amplitudes with definite $t$-channel naturality~\cite{lutz78}. Their tables are reproduced in the Appendix of Ref.~\cite{svec07a}. \\

In Tables I.-IV. we present expressions for the complete set of measurable $S$- and $P$-wave density matrix elements $\rho^\ell_k(ij)$ in terms of reduced transversity amplitudes $A$ and $\overline{A}$, $A=S,L,U,N$. The solution indices are suppressed for the sake of brevity. The expressions in the Tables I. and II. are valid for any phase $\omega$ and are not specific to the unique physical solution with $\omega=\pi$. The expressions in the Tables III. and IV. are valid only for the physical solution with $\omega=\pi$. The expressions yield a unique solution for final state density matrices $\rho_f(ij,\theta \phi,\vec{P})$. The mixed final state density matrix $\rho_f(\theta \phi,\vec{P})$ given by the Kraus representation (7.2) is thus unique as well. We conclude that Kraus representation of the mixed final state is uniquely determined by measurements on polarized target alone.

\section{Experimental determination of probabilities $p_{ij}$.}

A complete determination of the mixed final state $\rho_f(\theta \phi, \vec{P})$ requires knowledge of the probabilities $p_{ij}$ which are the diagonal elements of the initial state density matrix of the environment
\begin{equation}
\rho_i(E)=\sum \limits_{i,j=1,2} \sum \limits_{i',j'=1,2} p_{ij,i'j'}|ij><i'j'|
\end{equation}
In this Section we indicate how the probabilities $p_{ij} \equiv p_{ij,ij}$ can be determined in measurements of recoil hyperon polarization in processes such as $K^- p \to \pi^- \pi^+ \Lambda^0$ and $\pi^- p \to \pi- K^+ \Lambda^0$ using their self-analyzing weak decays.\\

Recoil nucleon (hyperon) polarization vector $\vec{Q} (ij,\theta \phi, \vec{P})$ is defined using a relation~\cite{svec07a}
\begin{equation}
\vec{Q} (ij,\theta \phi, \vec{P}) I^0(ij,\theta \phi, \vec{P}) \equiv \vec{I} (ij,\theta \phi, \vec{P})
\end{equation}
\noindent
We can write 
\begin{equation}
\rho_f(ij,\theta \phi, \vec{P}) = {1 \over{2}} (1+\vec{Q}(ij,\theta \phi, \vec{P}) \vec{\sigma}) I^0 (ij,\theta \phi, \vec{P})= \rho_n(\vec {Q(ij)}) I^0 (ij,\theta \phi, \vec{P}) 
\end{equation}
\noindent
The normalized final state density matrix $\rho^{'}_f(ij,\theta \phi, \vec{P}) =\rho_f(ij,\theta \phi, \vec{P})/{I^0(ij,\theta \phi, \vec{P})}$ is simply the spin density matrix of the recoil nucleon $\rho_n(\vec {Q}(ij))$. It will represent a pure final state if and only if the recoil nucleon polarization vector $\vec{Q} = (Q^1,Q^2,Q^3)$ satisfies the condition $|\vec{Q}|^2=1$ for all solid angles $\Omega = (\theta, \phi)$ at any given dipion mass $m$ and momentum transfer $t$~\cite{leader01,nielsen00}.\\

Because of the linearity of the mixed final state density matrix (7.2), it has a form similar to (7.6)
\begin{equation}
\rho_f(\theta \phi, \vec{P}) = {1 \over{2}} (I^0 (\theta \phi, \vec{P}) \sigma^0 + {\vec{I}} (\theta \phi, \vec{P}) \vec{\sigma})
\end{equation}
where
\begin{equation}
I^\ell(\theta \phi,\vec{P})=\sum \limits_{i,j=1,2} p_{ij}I^\ell(ij,\theta \phi, \vec{P})
\end{equation}
for $\ell=0,1,2,3$. The measured recoil nucleon polarization is similarly defined
\begin{equation}
\vec{Q} (\theta \phi, \vec{P}) I^0(\theta \phi, \vec{P}) \equiv \vec{I} (\theta \phi, \vec{P})
\end{equation}
\noindent
so that we can write the mixed state in the form
\begin{equation}
\rho_f(\theta \phi, \vec{P}) = {1 \over{2}} (1+\vec{Q}(\theta \phi, \vec{P}) \vec{\sigma}) I^0 (\theta \phi, \vec{P})= \rho_n(\vec {Q}) I^0 (\theta \phi, \vec{P}) 
\end{equation}
Inserting (8.3) into (7.2) and comparing with (8.7) we obtain
\begin{equation}
\vec{Q}(\theta \phi, \vec{P})=\sum \limits_{i,j=1,2}  \Bigl ( {p_{ij}I^0(ij,\theta \phi,\vec{P})\over{I^0(\theta \phi, \vec{P})}} \Bigr ) \vec{Q}(ij,\theta \phi, \vec{P})
\end{equation}
In the dipion mass region where $S$- and $P$ waves dominate the intensities $I^0(ij,\theta \phi, \vec{P})=I^0(\theta \phi, \vec{P})$ are the same for all $i,j=1,2$~\cite{svec07b} so that the measured recoil polarization is the average
\begin{equation}
\vec{Q}(\theta \phi, \vec{P})=\sum \limits_{i,j=1,2} p_{ij} \vec{Q}(ij,\theta \phi, \vec{P})
\end{equation}
Measurements on polarized targets determine uniquely the four sets of $S$- and $P$-wave density matrix elements and thus the four solutions of recoil polarizations $\vec{Q}(ij,\theta \phi, \vec{P})$ for any bin $(\theta \phi, m,t)$. Measurements of the recoil polarization $\vec{Q}(\theta \phi, \vec{P})$ in four distinct bins $(\theta \phi,m,t)$ provide four independent linear equations for probabilities $p_{ij}$ for each component of the polarization vector. The solutions for probabilities from all components must be consistent. The measurement of the probabilities $p_{ij}$ provides information about the diagonal elements of the density matrix of the environment $\rho_i(E)$ given by (8.1).\\

Recoil nucleon polarization is not experimentally accessible in $\pi^- p \to \pi^- \pi^+ n$. However, recoil hyperon polarization is accessible in measurements of $K^- p \to \pi^- \pi^+ \Lambda^0$ or $\pi^- p \to \pi^- K^+ \Lambda^0$ through weak decays $\Lambda^0 \to p \pi^-$ in measurements on unpolarized or polarized targets. Using the spin formalism for two-particle decays of a particle with spin ${1 \over{2}}$ developed in Sections 8.2.1.(i) and 8.2.1.(v) of the monograph on spin physics by E.~Leader~\cite{leader01}, it is straightforward to derive a general form of normalized angular distribution $W(\theta_p, \phi_p)$ of the protons in $\Lambda^0$ decays. It reads 
\begin{equation}
W(\theta_p,\phi_p)={1\over{4 \pi}} \Bigl ( 1+\alpha Q^1\sin\theta_p \cos \phi_p+\alpha Q^2 \sin \theta_p \sin \phi_p +\alpha Q^3 \cos \theta_p \Bigr )
\end{equation}
where $\theta_p, \phi_p$ is the direction of the decay proton in center-of-mass system. $\alpha$ is a weak decay parameter measuring the real part of the interference between parity conserving and parity violating components of the decay amplitude and it is well known experimentally~\cite{leader01}.\\

With sufficiently high statistics in $(\theta \phi,m,t)$ bins there will be sufficient statistics of $\Lambda^0$ decays to determine the recoil polarization vector $\vec{Q}(\theta \phi,\vec{P})$ with high precission. Simultaneous amplitude analysis of $K^- p \to \pi^- \pi^+ \Lambda^0$ or $\pi^- p \to \pi^- K^+ \Lambda^0$ on polarized target will determine the four solutions for the recoil polarizations $\vec{Q}(ij,\theta \phi,\vec{P})$, presuming the phase $\omega=\pi$ is still the physical solution. In principle, these self-analyzing processes can thus determine the probabilities $p_{ij}$.

\section{Conclusions.}

We have presented the first model independent determination of unnatural exchange $S$- and $P$-wave helicity amplitudes in $\pi^-p \to \pi^- \pi^+ n$ from CERN measurements on transversely polarized target at 17.2 GeV/c for dipion masses in the range 580-1080 MeV and momentum transfers $0.005 \leq -t \leq 0.20$ (GeV/c)$^2$. The analytical determination of helicity amplitudes is made possible by our finding of analytical solutions for the relative phase $\omega_{ij}=\Phi_{S_d}(j)-\Phi_{S_u}(i)$ from the four sets of reduced transversity amplitudes $A(i),\overline{A}(j),i,j=1,2$, $A=S,L,U$ found also analytically from the measured data~\cite{svec07b}.\\

The solutions for helicity amplitudes corresponding to two solutions for $\omega$ with $\cos \omega \neq 0$ and $\sin \omega \neq 0$ show chaotic behaviour and lack a clear resonant structure at $\rho^0(770)$ mass, and are rejected. Three solutions for helicity amplitudes corresponding to solution for $\omega$ with $\sin \omega = \pm 1$ and $\cos \omega=+1$ are also rejected since they do not satisfy the requirement of pion exchange dominance of helicity flip amplitudes. This leaves a unique physical solution for helicity amplitudes corresponding to $\cos \omega =-1$, or $\omega_{ij}= \pi$. The solution is unique up to the sign ambiguity of phases of reduced transversity amplitudes to be resolved unambigously by measurements with planar target polarization~\cite{svec07b}.\\

Assigning $\rho^0(770)$ phase to the dominant $P$-wave helicity flip amplitude $L_1(ij)$ necessitates a phase of the $S$-wave helicity flip amplitude $S_1(ij)$ that is near to the $\rho^0(770)$ phase. Both amplitudes show resonant structures around 980 MeV for all $i,j=1,2$. These amplitudes are thus consistent with $\rho^0(770)-f_0(980)$ mixing observed previously in the reduced transversity amplitudes~\cite{svec07b}.\\

The physical solutions for the relative phases $\omega_{ij}=\pi$ satisfy trivially the self-consistency condition (2.7) that must be satisfied by co-evolution amplitudes (3.8). The four sets of physical solutions for transversity amplitudes $A_u(i),A_d(j),i,j=1,2$ can thus be identified with the four sets of co-evolution amplitudes (3.8) connecting in a unique way the Kraus representation to the experimentally measured amplitudes. This connection validates further the view of pion creation process $\pi^- p \to \pi^- \pi^+ n$ as an open quantum system interacting with a quantum environment~\cite{svec07a}.  Information about the quantum state of the environment is accessible in part by determination of the probabilities $p_{ij}$ in measurements of recoil hyperon polarization in reactions such as $\pi^- p \to \pi^- K^+ \Lambda^0$ and $K^- p \to \pi^- \pi^+ \Lambda^0$ on polarized target.\\

{\bf Availability of data files from amplitude analyses.}\\

Output data files from computer codes for amplitude analyses are available on request to the author at svec@hep.physics.mcgill.ca. Code AACERNM performs amplitude analysis of mass dependence of the CERN data for polarized or unpolarized target. It calculates moduli and phases of reduced transversity amplitudes, partial wave intensities and polarizations, interference terms, numerous tests and auxiliary calculations. The output file for polarized target is AACERNM1p. The moduli and phases of reduced transversity amplitudes form input file for computer code HACERNMa. The code HACERNMa selects the solution $n$ for $\omega$ and calculates first moduli squared and relative phases of helicity amplitudes and performs tests on their solutions before it calculates moduli and phases as well as real and imaginary parts of both transversity and helicity amplitudes in output file HACERNMa$n$.

\begin{figure}
\includegraphics[width=16cm,height=21.5cm]{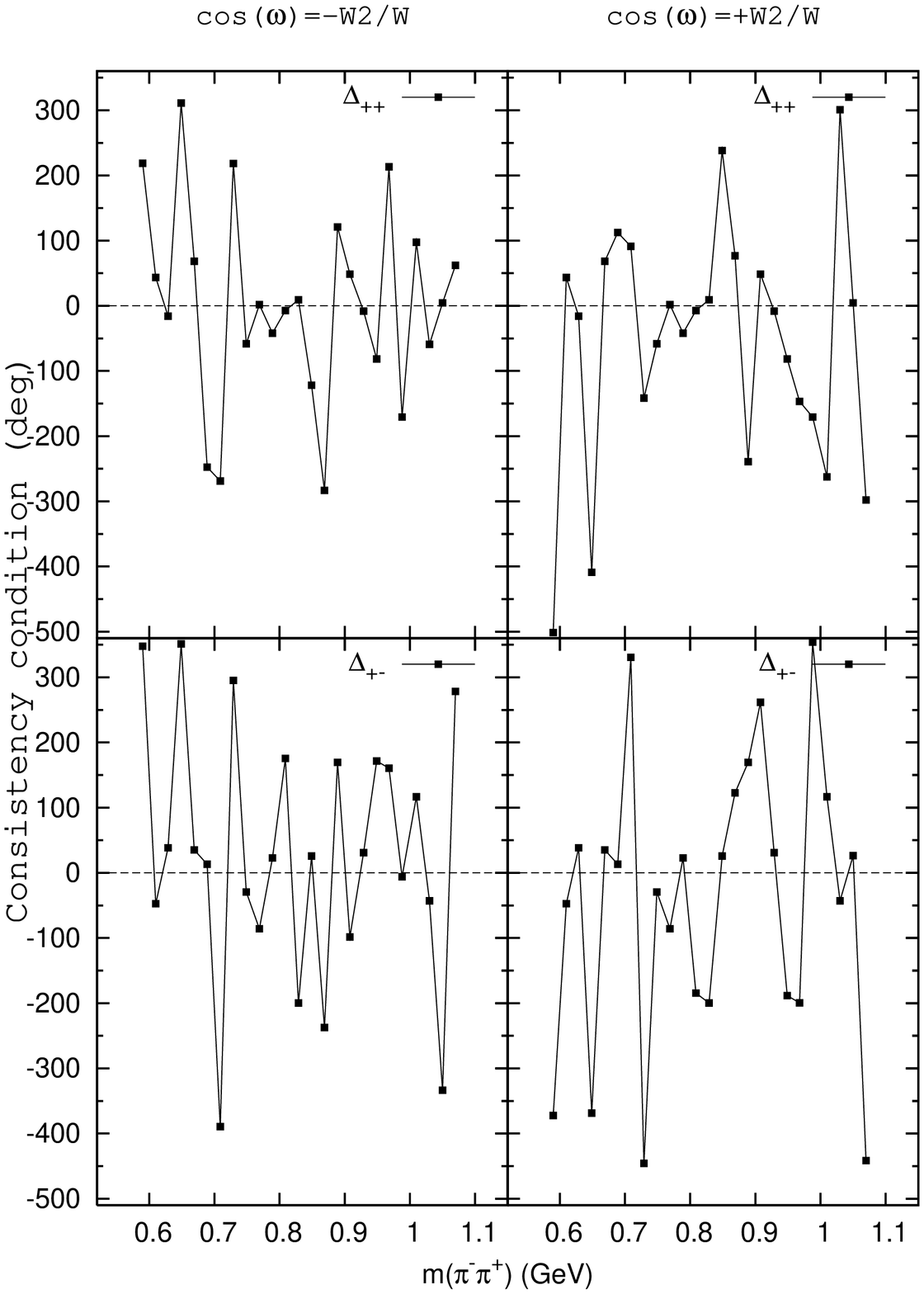}
\caption{Test of consistency of the four sets of solutions $A_u(i),A_d(j),i,j=1,2$ with $\cos \omega = \pm W_2/W$. The assumption that all four sets are physical requires that $\Delta=(\omega_{11}-\omega_{21})-(\omega_{12}-\omega_{22}) \equiv 0.$}
\label{Figure 1}
\end{figure}

\begin{figure}
\includegraphics[width=16cm,height=21.5cm]{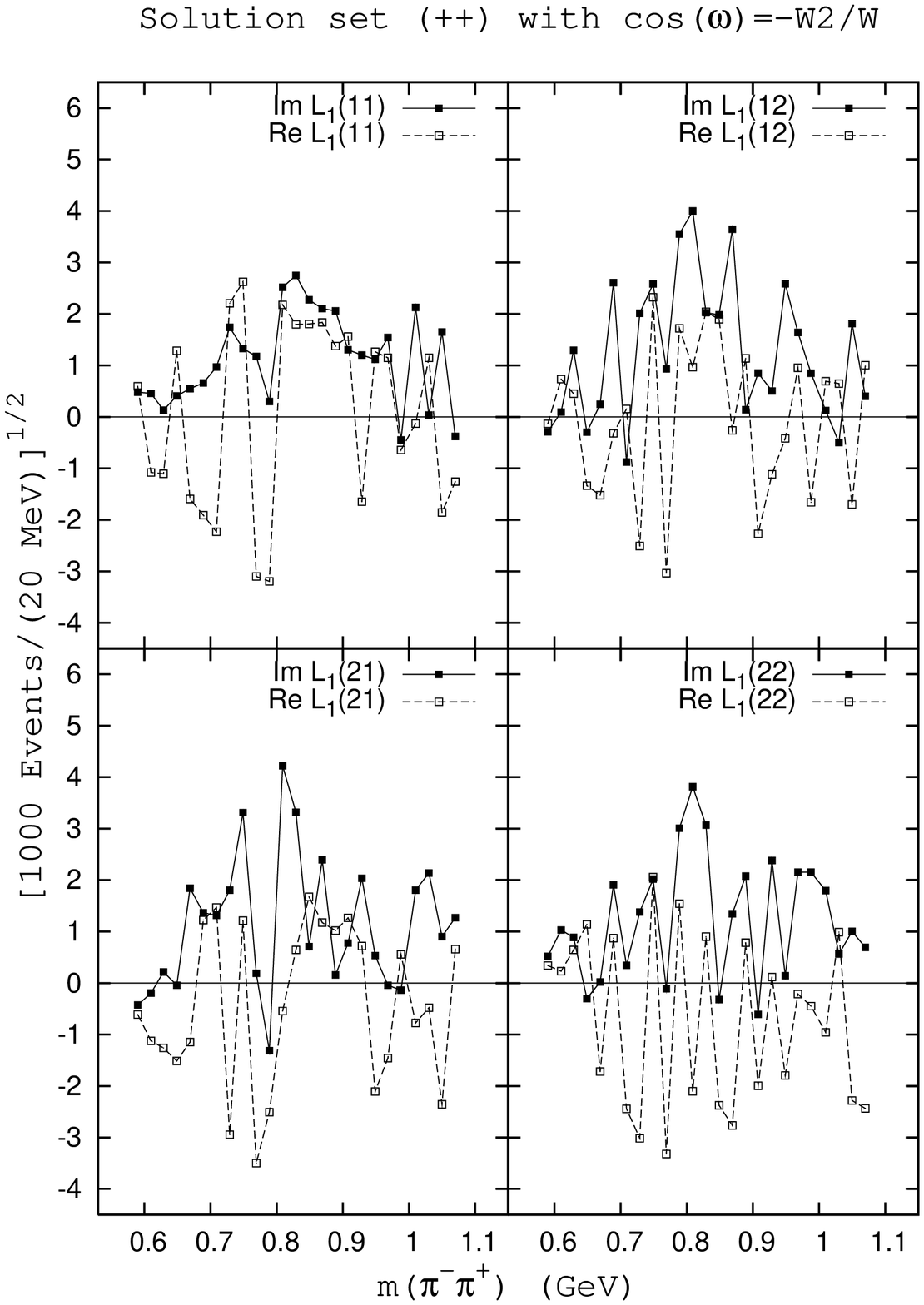}
\caption{Solutions for helicity flip amplitude $L_1$ with signs of phases $++$ and $\cos \omega = - W_2/W$.}
\label{Figure 2}
\end{figure}

\begin{figure}
\includegraphics[width=16cm,height=21.5cm]{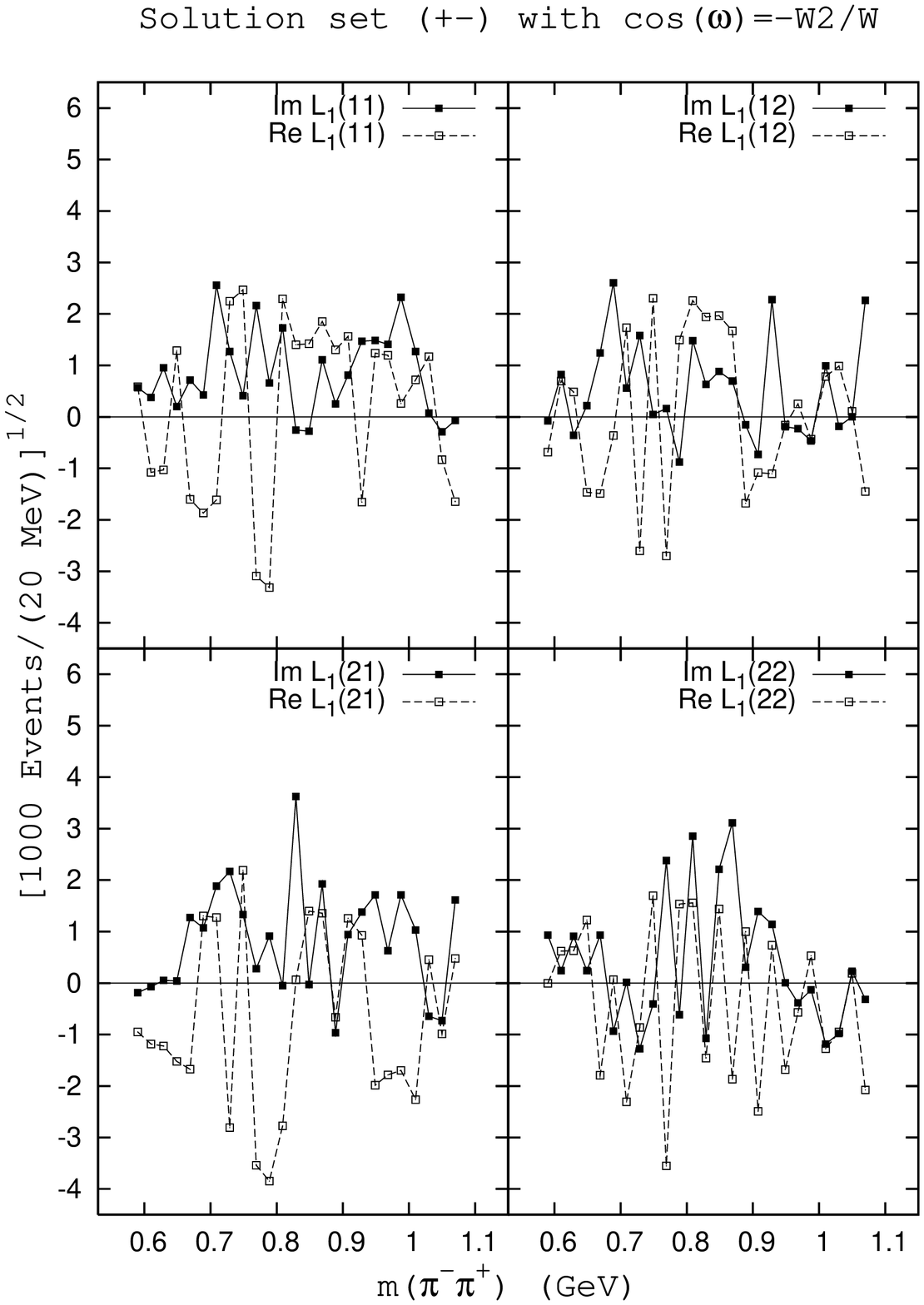}
\caption{Solutions for helicity flip amplitude $L_1$ with signs of phases $+-$ and $\cos \omega = - W_2/W$.}
\label{Figure 3}
\end{figure}

\begin{figure}
\includegraphics[width=16cm,height=21.5cm]{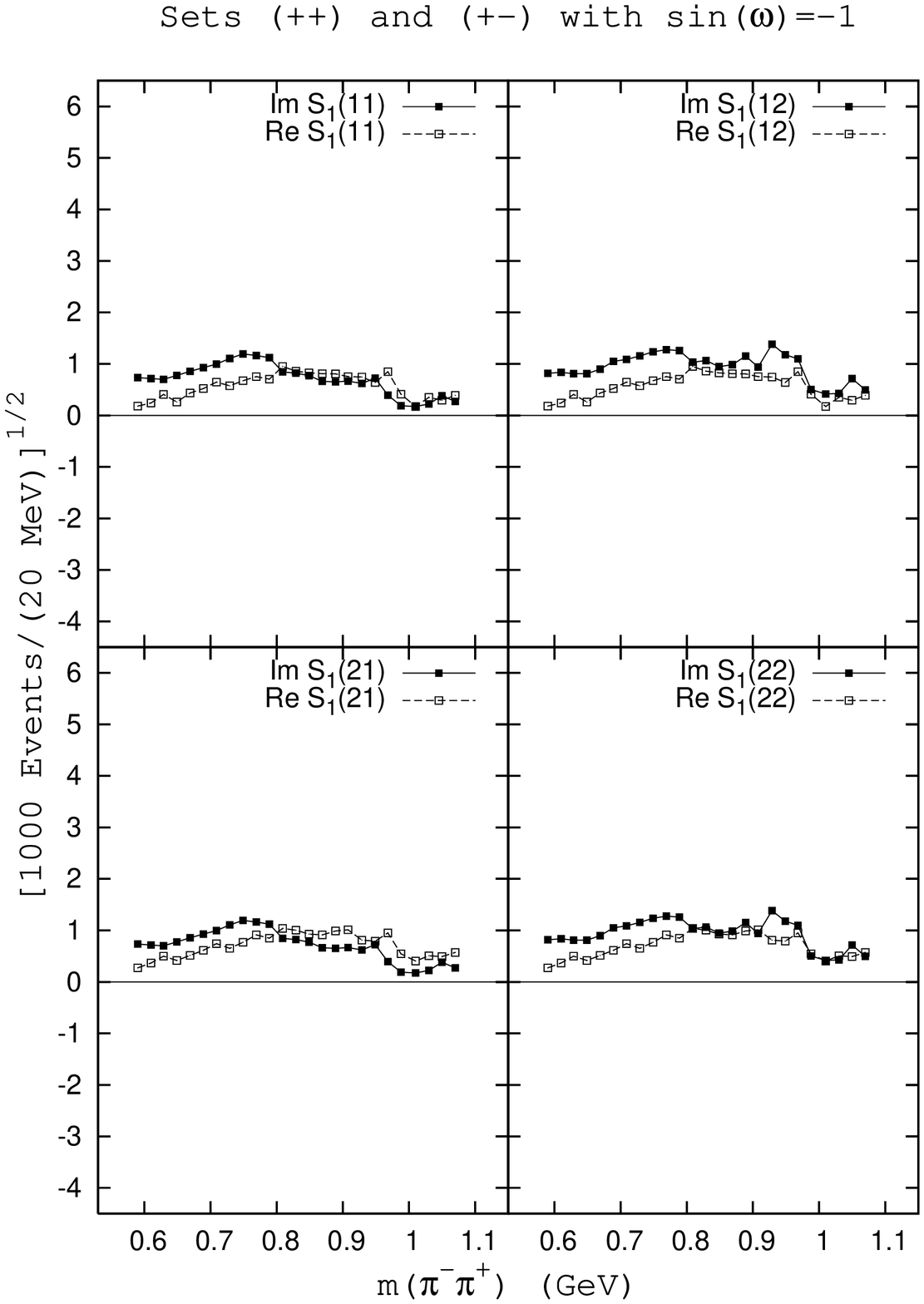}
\caption{Solutions for helicity flip amplitude $S_1$ with $\sin \omega =-1$. The solutions are the same bor both sets $++$ and $+-$ of signs of phases. The helicity flip amplitude $S_1={1\over{\sqrt{2}}}(|S|+i|\overline{S}|)$ and helicity non-flip amplitude $S_0={1\over{\sqrt{2}}}(|\overline{S}|+i|S|)$.}
\label{Figure 4}
\end{figure}

\begin{figure}
\includegraphics[width=16cm,height=21.5cm]{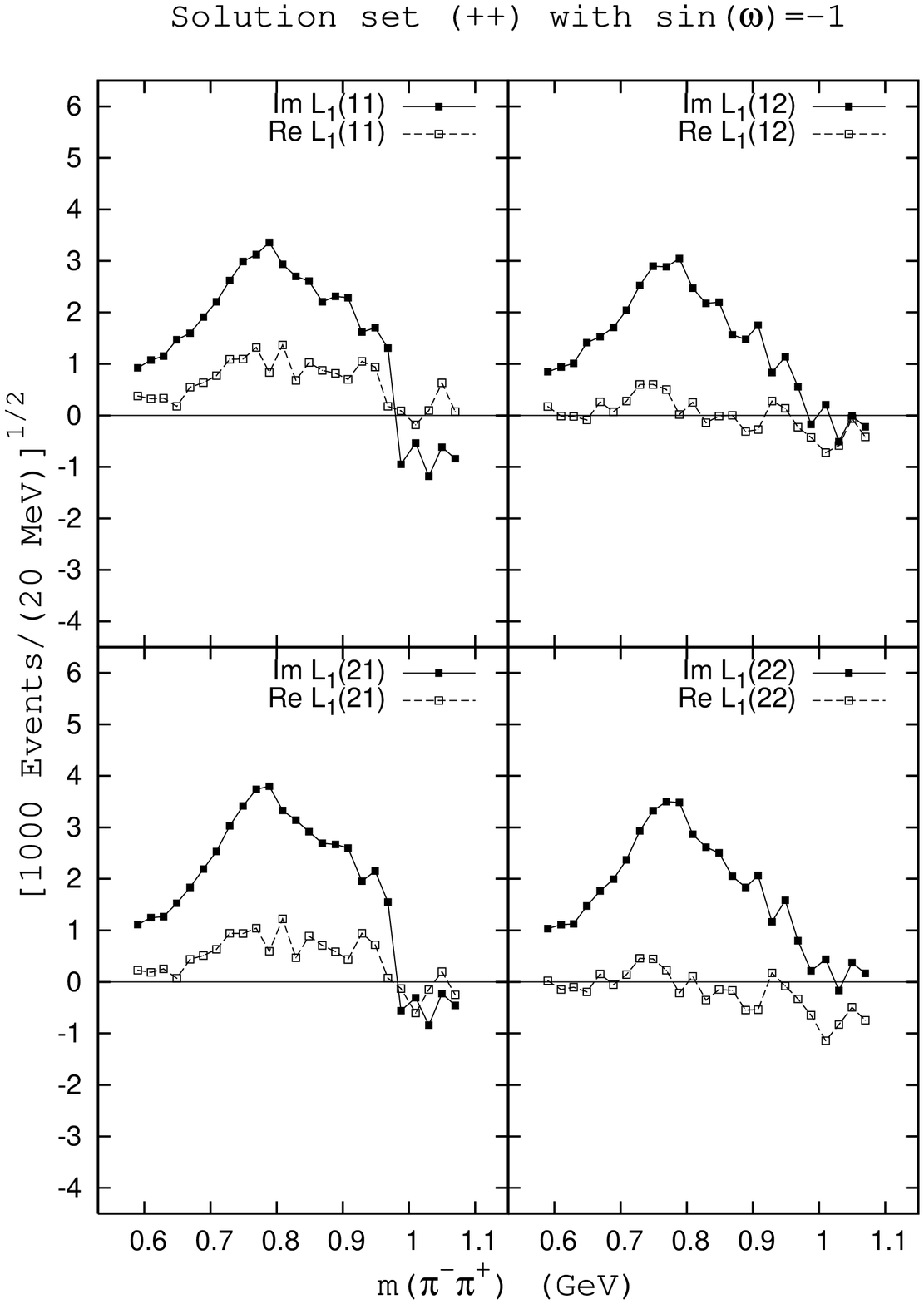}
\caption{Solutions for helicity flip amplitude $L_1$ with signs of phases $++$ and $\sin \omega = -1$.}
\label{Figure 5}
\end{figure}

\begin{figure}
\includegraphics[width=16cm,height=21.5cm]{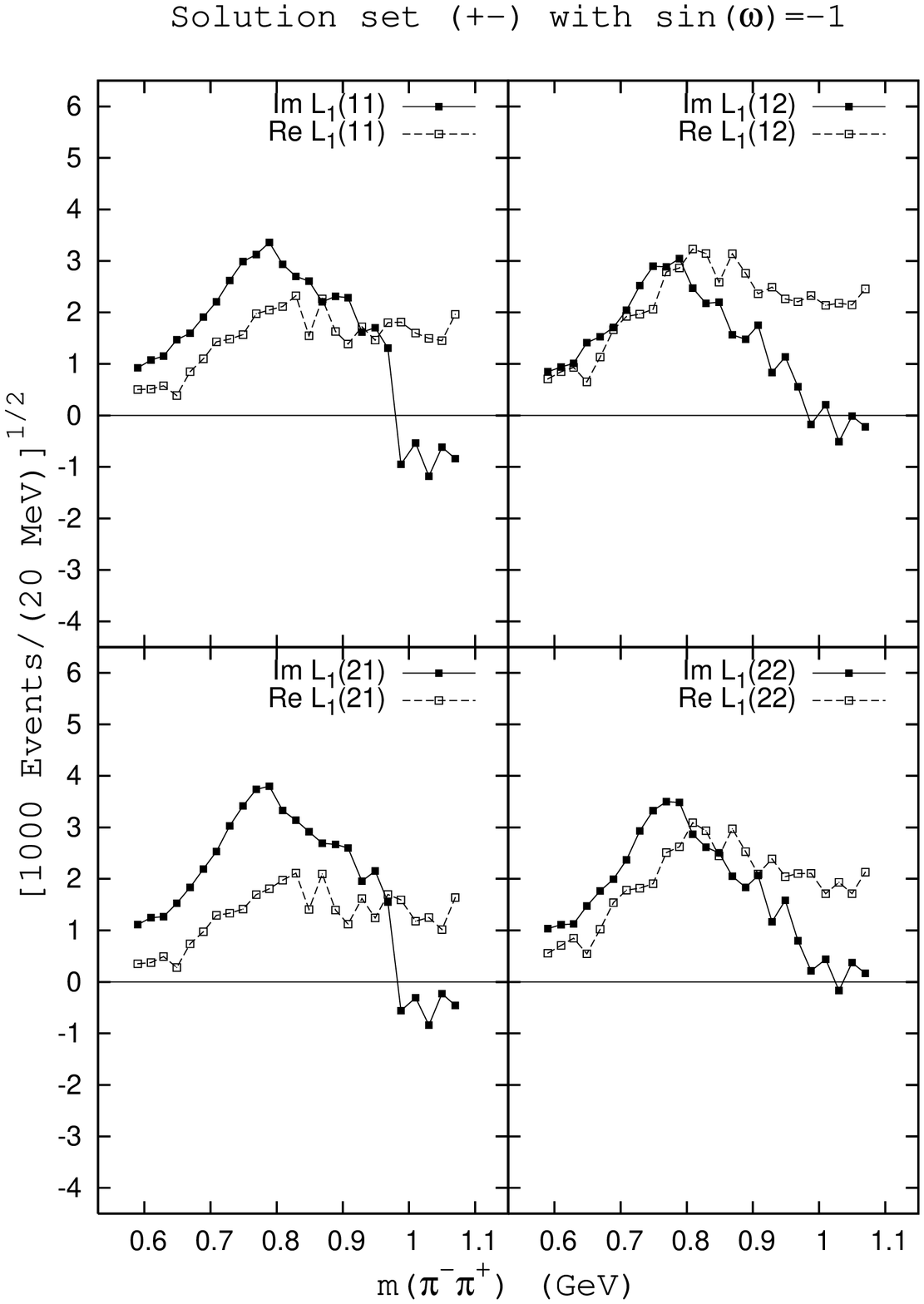}
\caption{Solutions for helicity flip amplitude $L_1$ with signs of phases $+-$ and $\sin \omega = -1$.}
\label{Figure 6}
\end{figure}

\begin{figure}
\includegraphics[width=16cm,height=21.5cm]{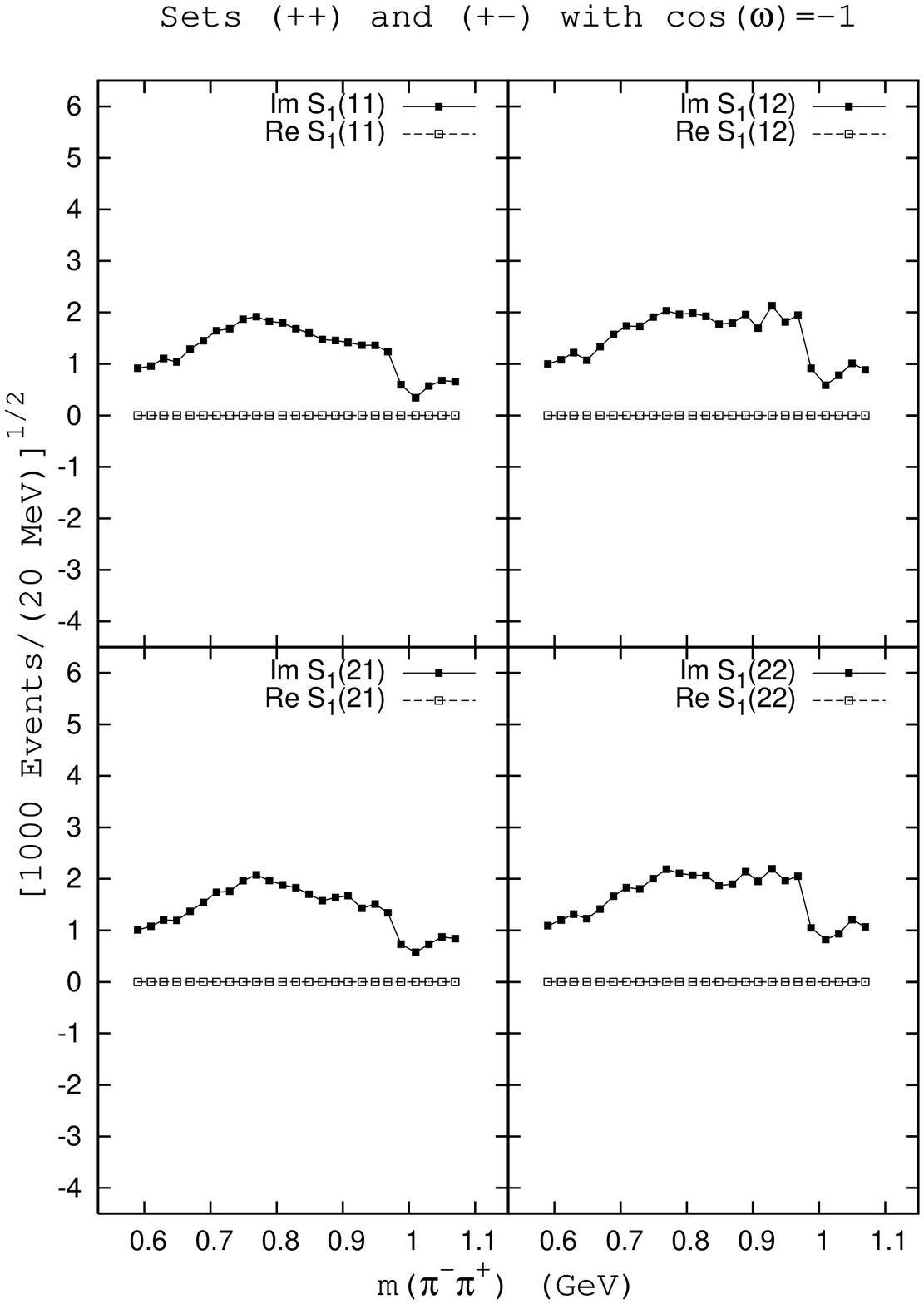}
\caption{Solutions for helicity flip amplitude $S_1$ with $\cos \omega =-1$. The solutions are the same bor both sets $++$ and $+-$ of signs of phases. The helicity flip amplitude $S_1={i\over{\sqrt{2}}}(|\overline{S}|+|S|)$ and helicity non-flip amplitude $S_0={1\over{\sqrt{2}}}(|\overline{S}|-|S|)$ (Fig. 12).}
\label{Figure 7}
\end{figure}

\begin{figure}
\includegraphics[width=16cm,height=21.5cm]{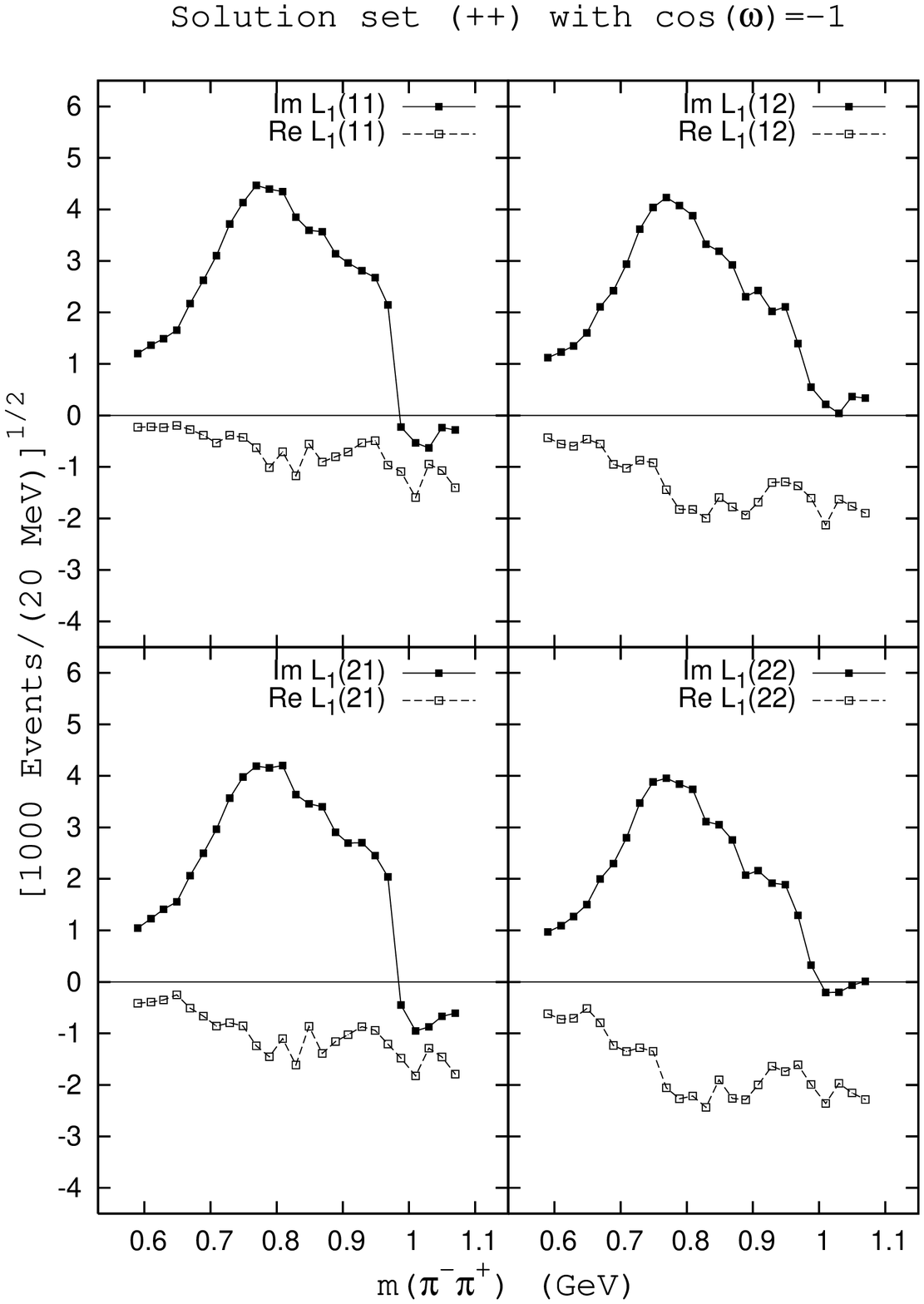}
\caption{Solutions for helicity flip amplitude $L_1$ with signs of phases $++$ and $\cos \omega = -1$.}
\label{Figure 8}
\end{figure}

\begin{figure}
\includegraphics[width=16cm,height=21.5cm]{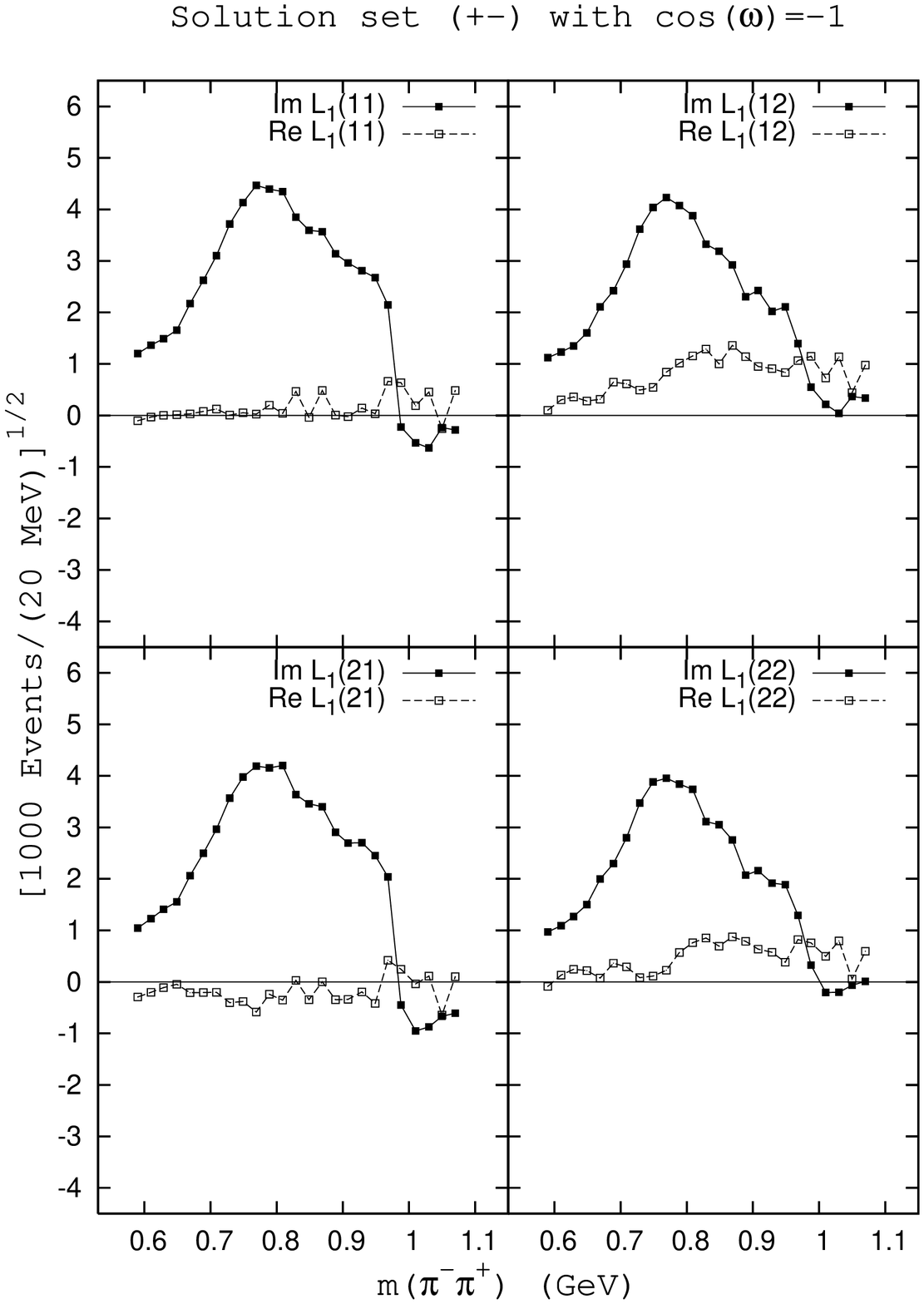}
\caption{Solutions for helicity flip amplitude $L_1$ with signs of phases $+-$ and $\cos \omega = -1$.}
\label{Figure 9}
\end{figure}

\begin{figure}
\includegraphics[width=16cm,height=21.5cm]{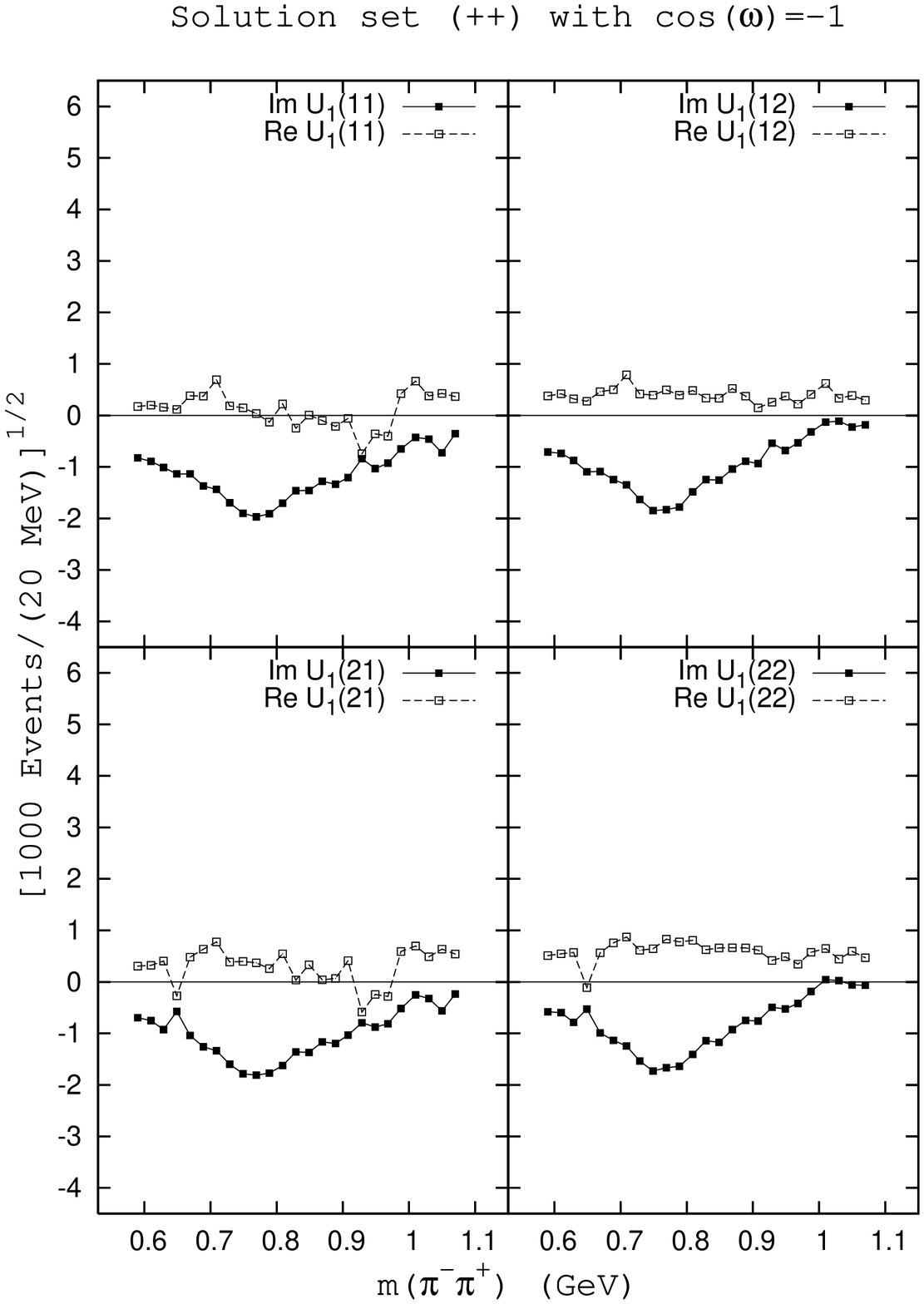}
\caption{Solutions for helicity flip amplitude $U_1$ with signs of phases $++$ and $\cos \omega = -1$.}
\label{Figure 10}
\end{figure}

\begin{figure}
\includegraphics[width=16cm,height=21.5cm]{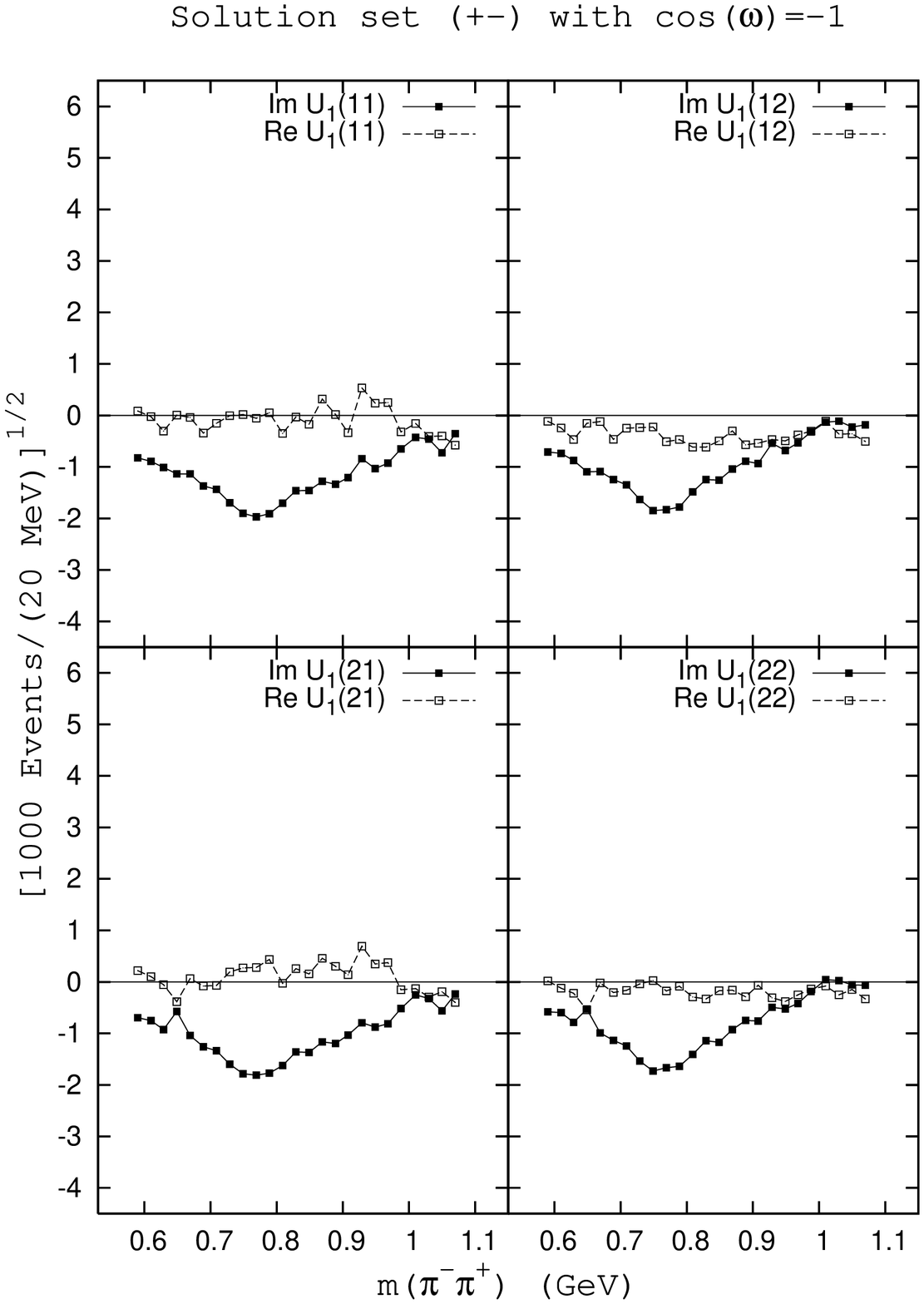}
\caption{Solutions for helicity flip amplitude $U_1$ with signs of phases $+-$ and $\cos \omega = -1$.}
\label{Figure 11}
\end{figure}

\begin{figure}
\includegraphics[width=16cm,height=21.5cm]{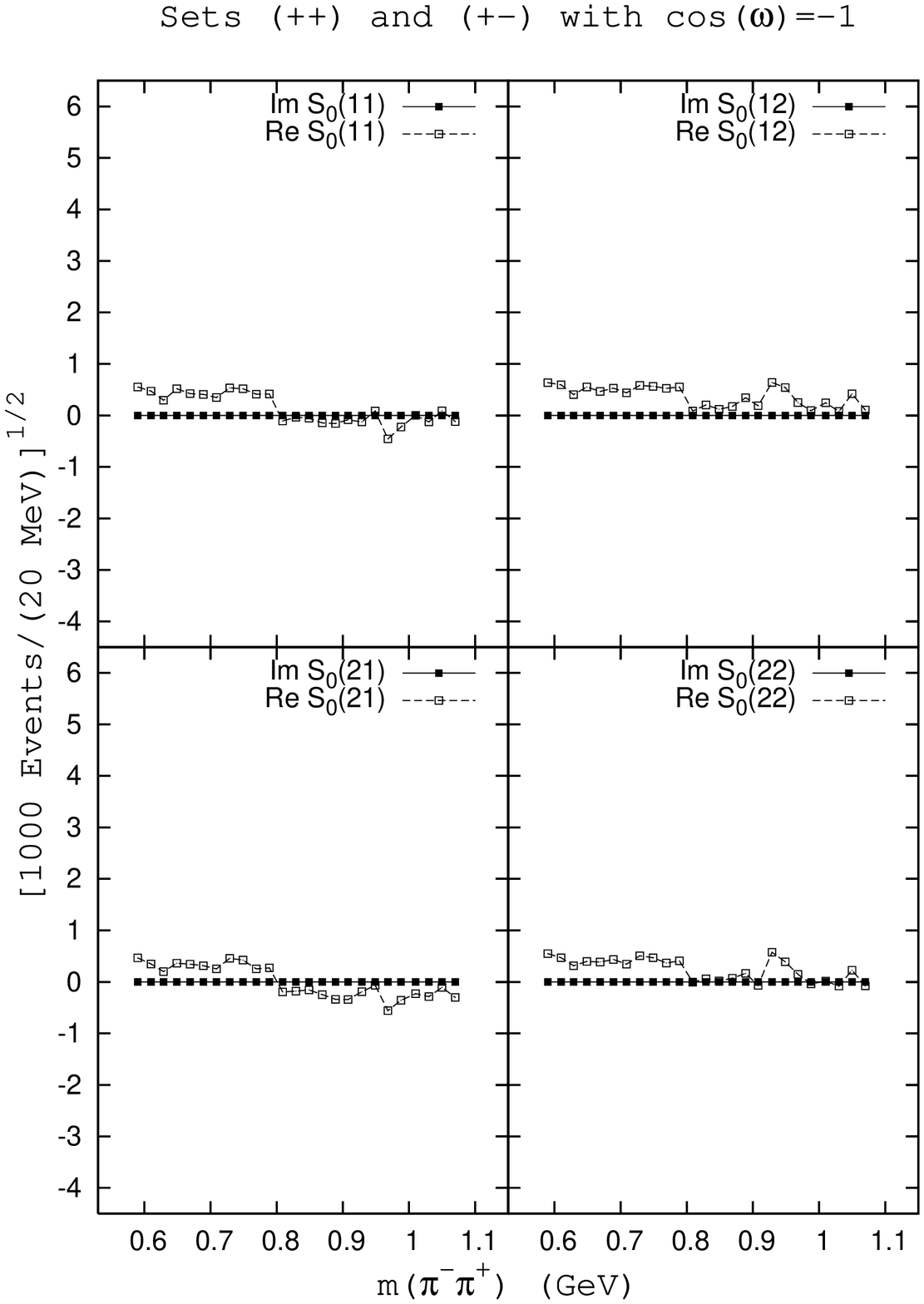}
\caption{Solutions for helicity non-flip amplitude $S_0$ with $\cos \omega =-1$. The solutions are the same bor both sets $++$ and $+-$ of signs of phases. The helicity non-flip amplitude $S_0={1\over{\sqrt{2}}}(|\overline{S}|-|S|)$ and helicity flip amplitude $S_1={i\over{\sqrt{2}}}(|\overline{S}|+|S|)$ (Fig. 7).}
\label{Figure 12}
\end{figure}

\begin{figure}
\includegraphics[width=16cm,height=21.5cm]{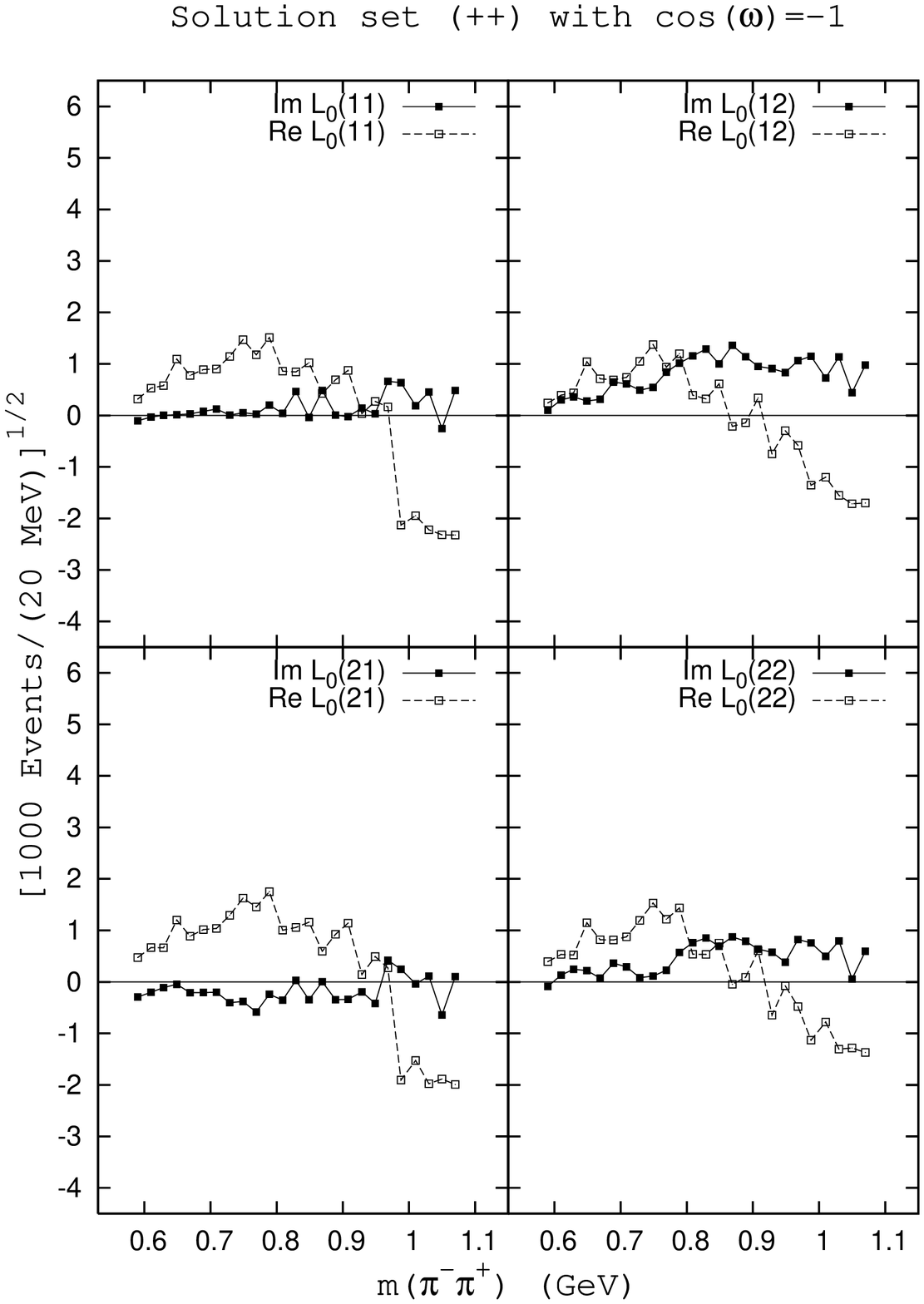}
\caption{Solutions for helicity non-flip amplitude $L_0$ with signs of phases $++$ and $\cos \omega = -1$.}
\label{Figure 13}
\end{figure}

\begin{figure}
\includegraphics[width=16cm,height=21.5cm]{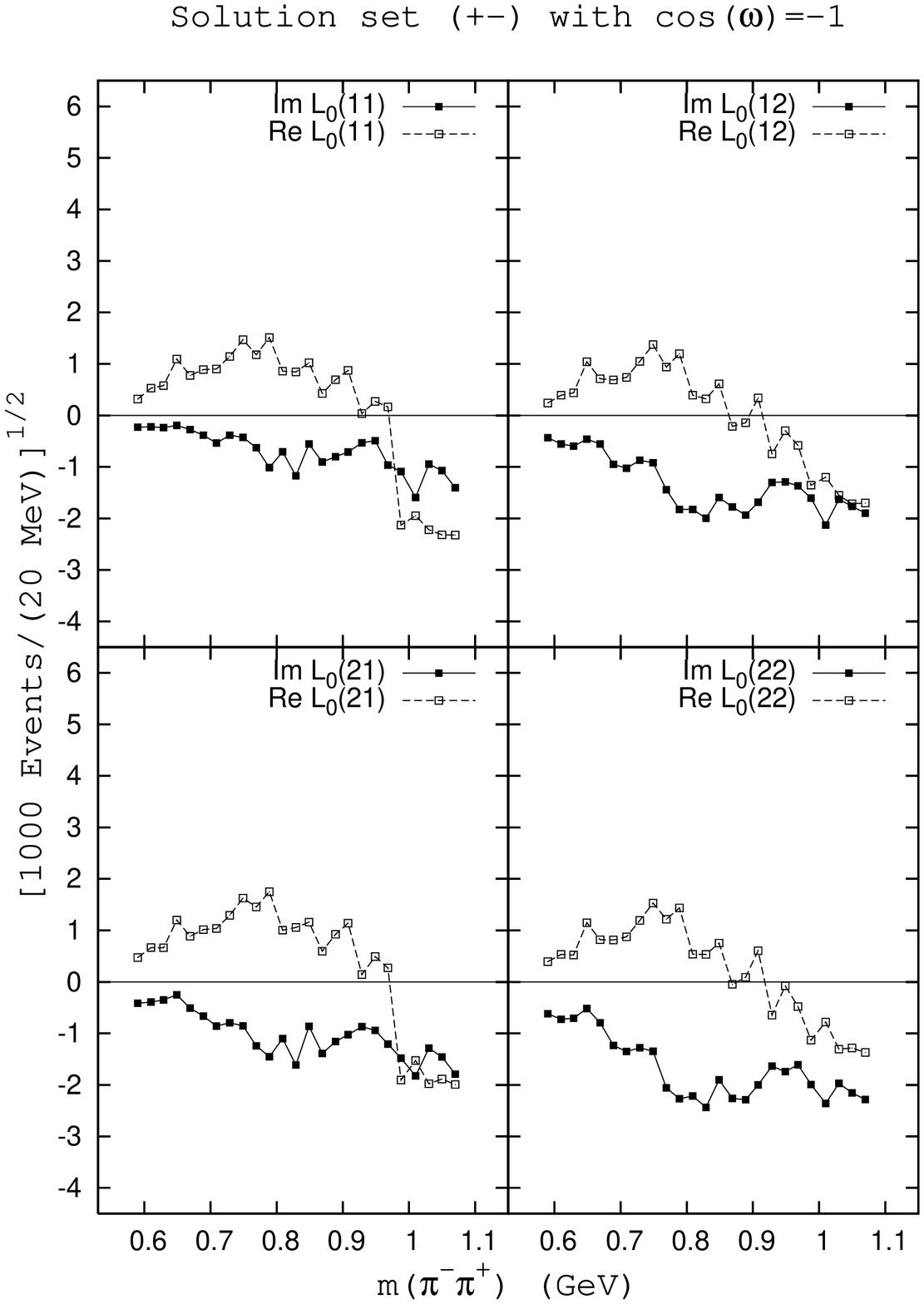}
\caption{Solutions for helicity non-flip amplitude $L_0$ with signs of phases $+-$ and $\cos \omega = -1$.}
\label{Figure 14}
\end{figure}

\begin{figure}
\includegraphics[width=16cm,height=21.5cm]{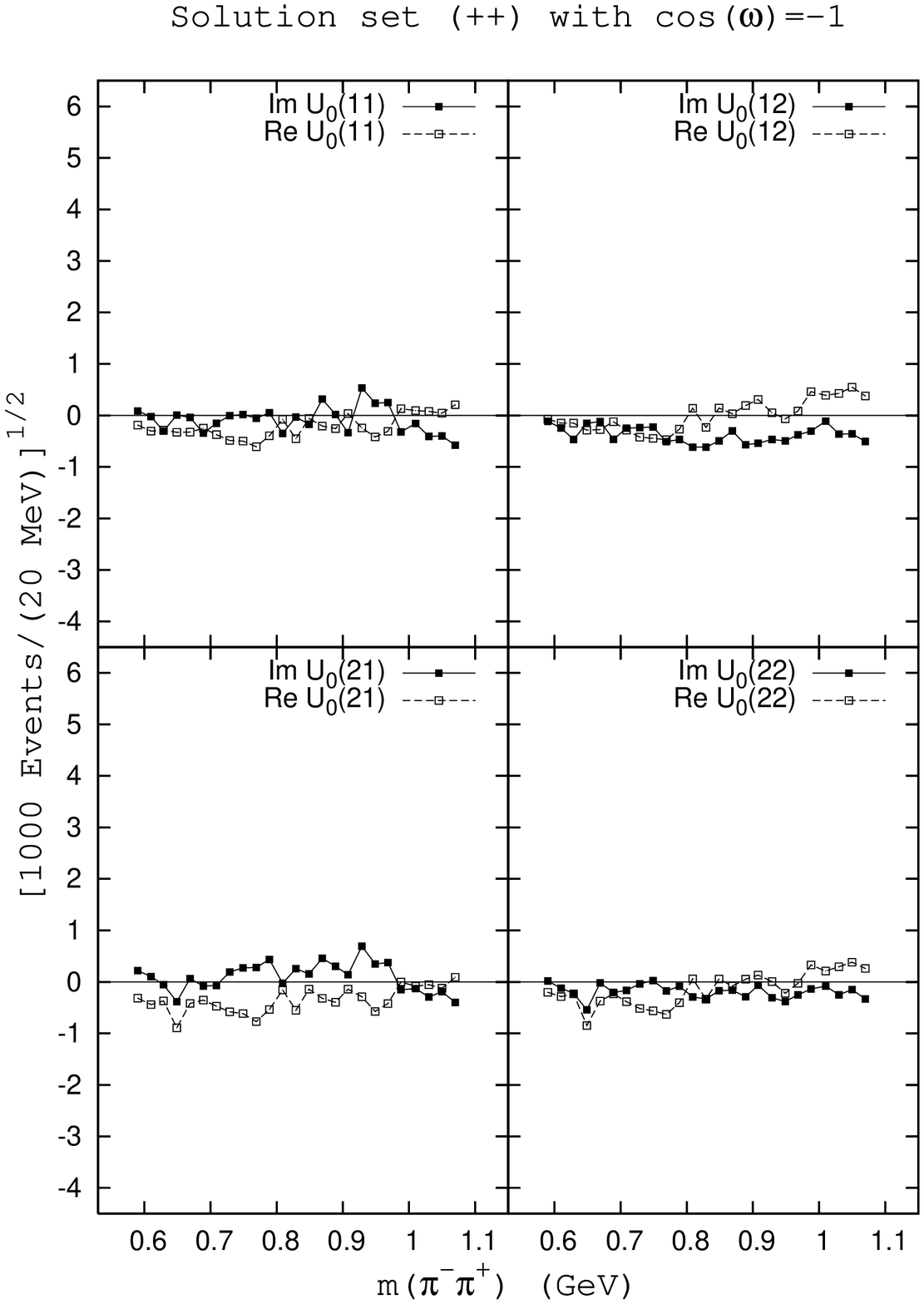}
\caption{Solutions for helicity non-flip amplitude $U_0$ with signs of phases $++$ and $\cos \omega = -1$.}
\label{Figure 15}
\end{figure}

\begin{figure}
\includegraphics[width=16cm,height=21.5cm]{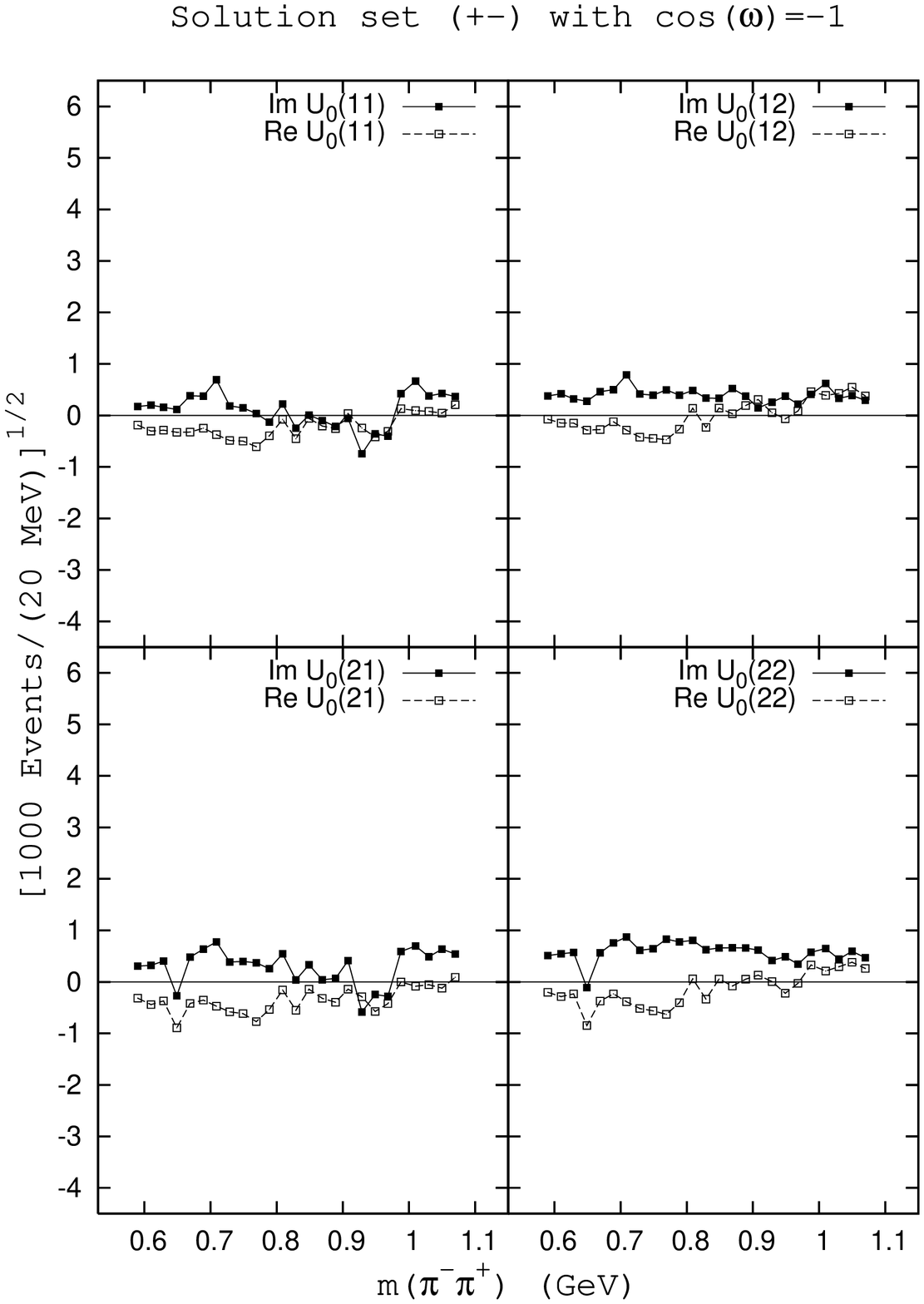}
\caption{Solutions for helicity non-flip amplitude $U_0$ with signs of phases $+-$ and $\cos \omega = -1$.}
\label{Figure 16}
\end{figure}

\begin{figure}
\includegraphics[width=16cm,height=21.5cm]{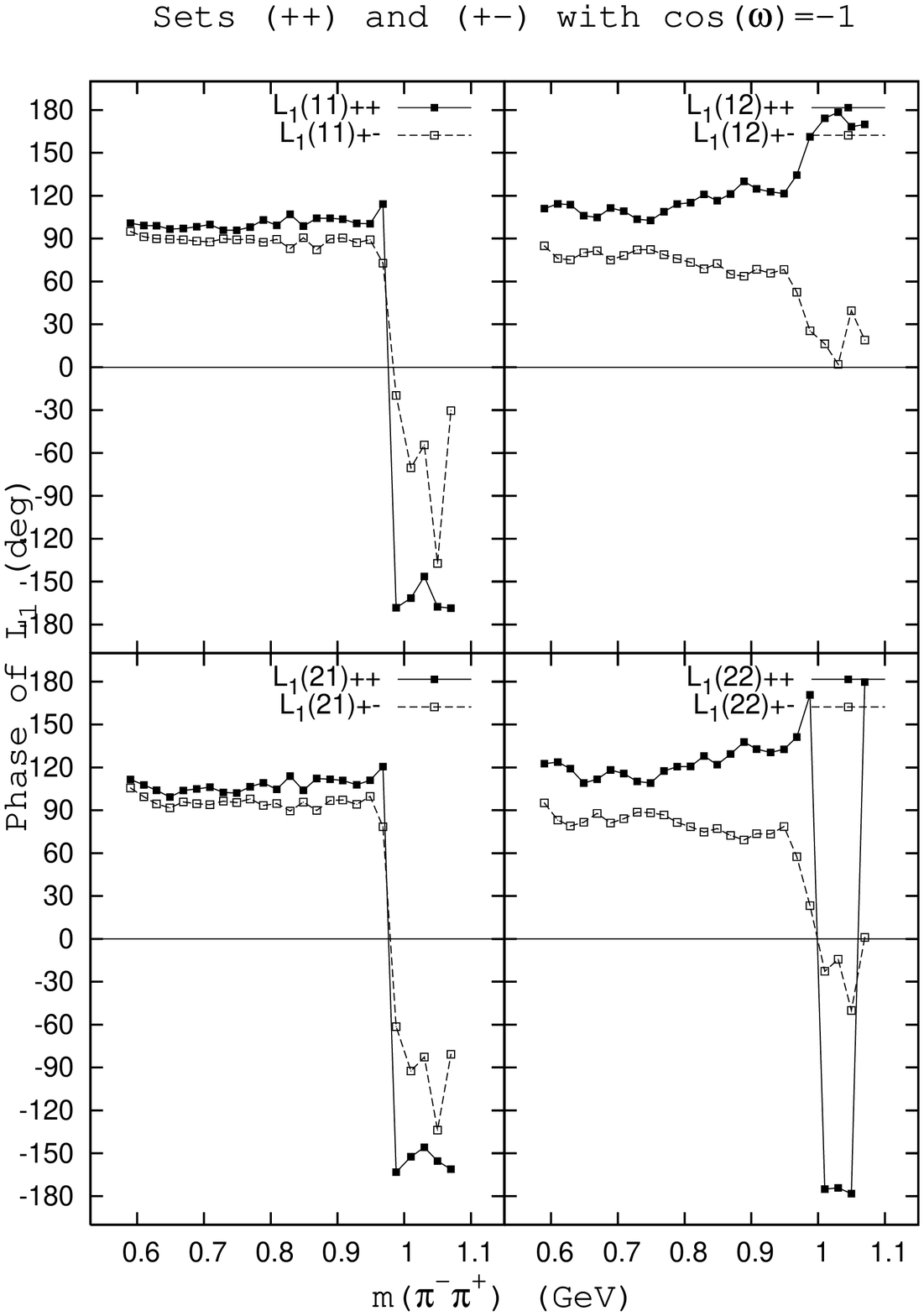}
\caption{Original phases of helicity flip amplitude $L_1$ for solutions with $\cos \omega = -1$. Absolute phase is set at $\Phi_{S_u}(i)=-\pi,i=1,2$.}
\label{Figure 17}
\end{figure}

\begin{figure}
\includegraphics[width=16cm,height=21.5cm]{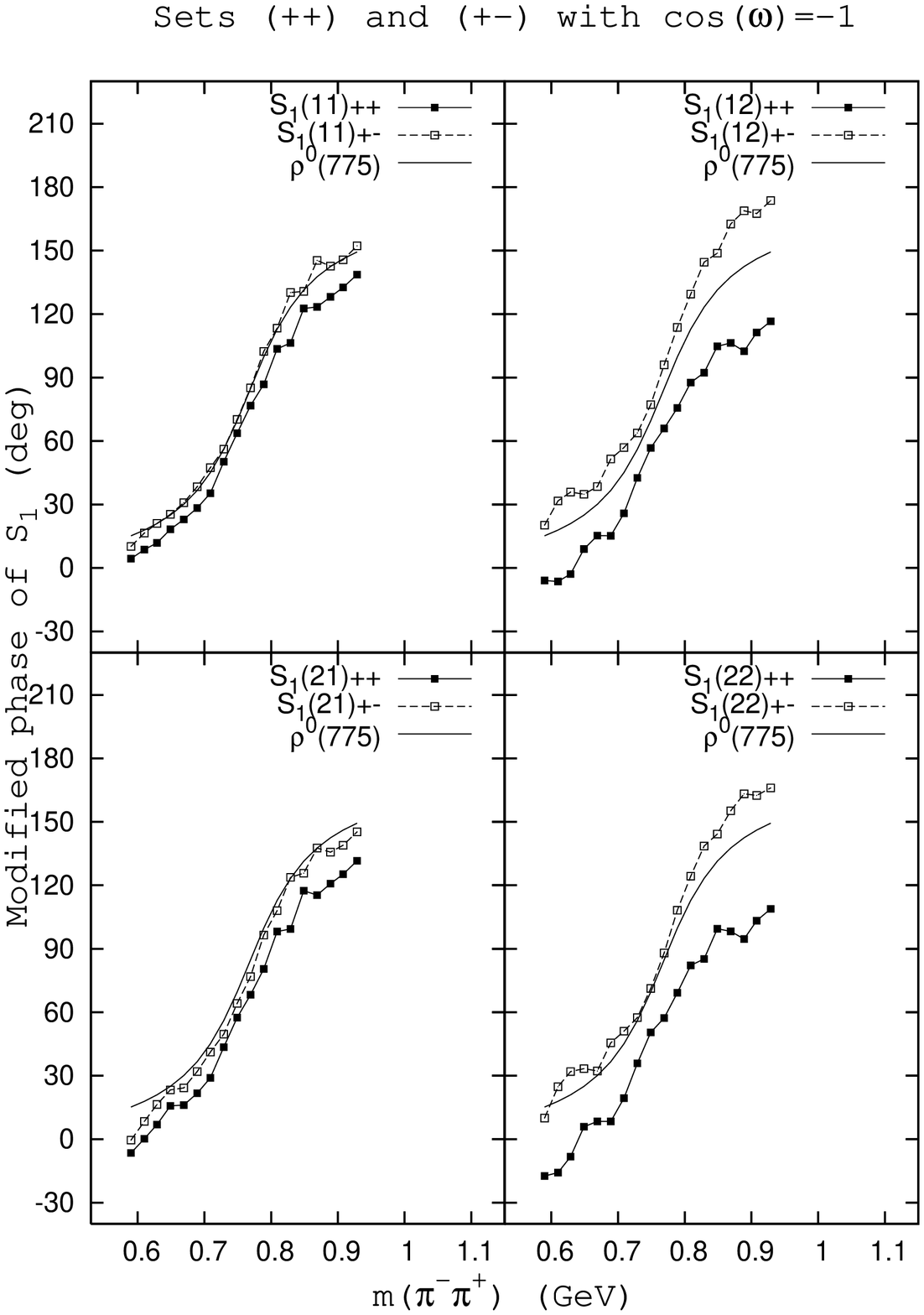}
\caption{Modified phases of helicity non-flip amplitude $S_0$ for solutions with $\cos \omega = -1$ after assigning the amplitude $L_1$ the Breit-Wigner phase $\Phi(\rho^0)$ of $\rho^0(770)$ resonance.}
\label{Figure 18}
\end{figure}

\begin{figure}
\includegraphics[width=16cm,height=21.5cm]{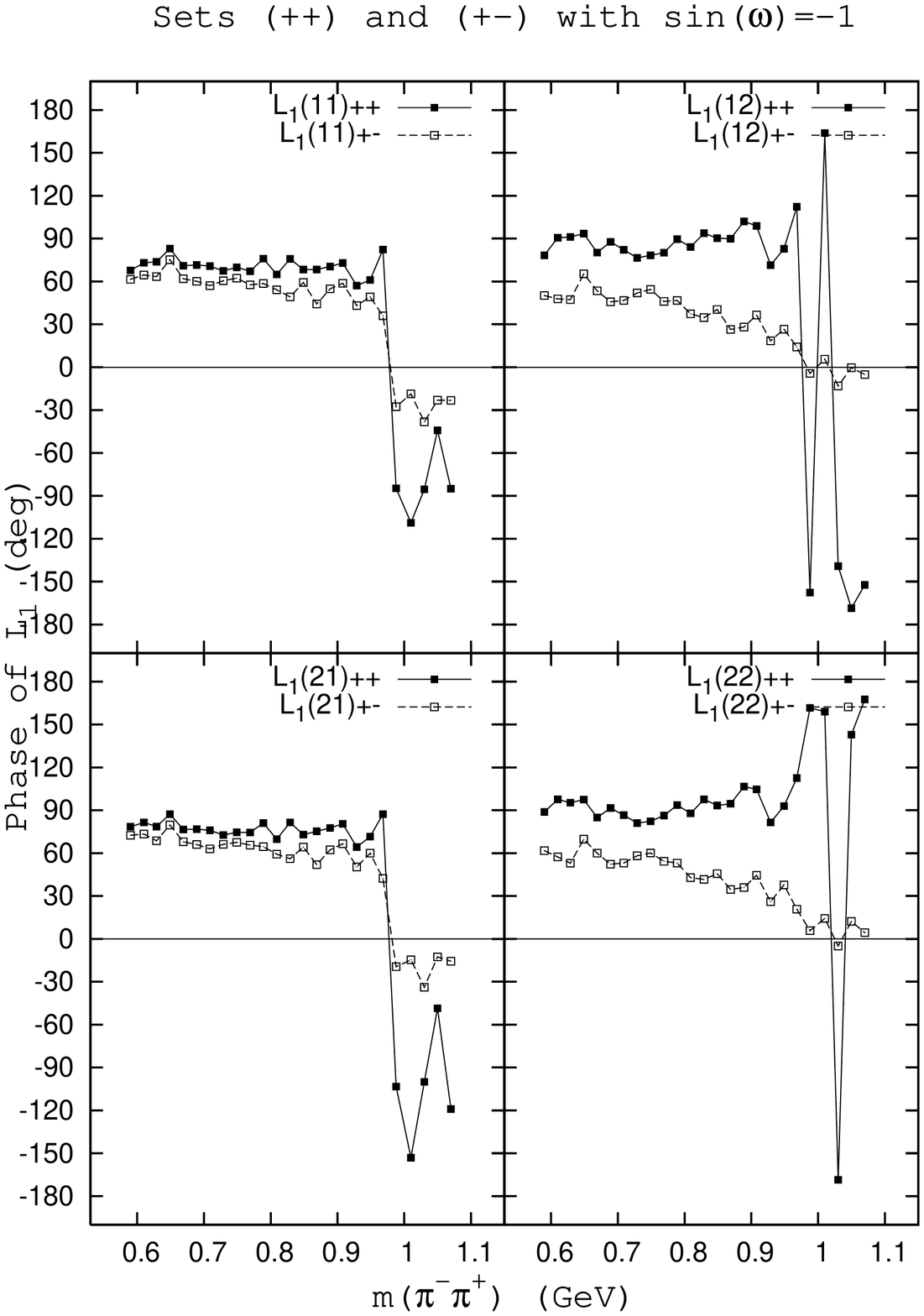}
\caption{Original phases of helicity flip amplitude $L_1$ for solutions with $\sin \omega = -1$. Absolute phase is set at $\Phi_{S_u}(i)=+\pi/2,i=1,2$.}
\label{Figure 19}
\end{figure}

\begin{figure}
\includegraphics[width=16cm,height=21.5cm]{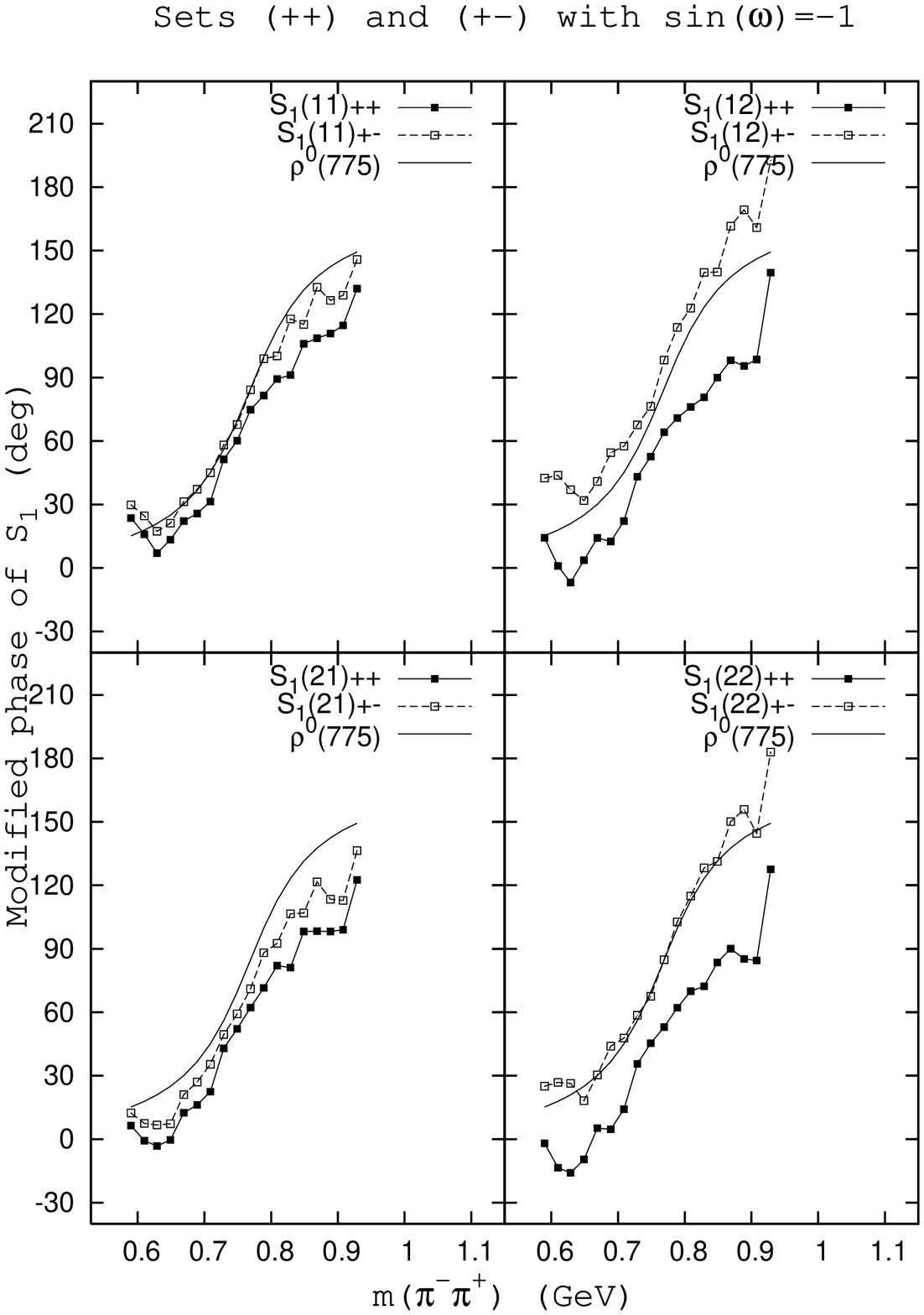}
\caption{Modified phases of helicity non-flip amplitude $S_0$ for solutions with $\sin \omega = -1$ after assigning the amplitude $L_1$ the Breit-Wigner phase $\Phi(\rho^0)$ of $\rho^0(770)$ resonance.}
\label{Figure 20}
\end{figure}

\end{document}